\documentclass[fleqn,usenatbib]{mnras}
\usepackage{txfonts} 
\usepackage[T1]{fontenc}
\usepackage{graphicx}

\title{The Gaia DR1 Mass-Radius Relation for White Dwarfs}

\author[Tremblay et al.]  {P.-E. Tremblay$^{1}$\thanks{E-mail:
    P-E.Tremblay@warwick.ac.uk}, N. Gentile-Fusillo$^{1}$, R. Raddi$^{1}$,
  S. Jordan$^{2}$, C. Besson$^{3}$,
  \newauthor{B.~T. G\"ansicke$^{1}$, S.~G. Parsons$^{4}$, D. Koester$^{5}$,
    T. Marsh$^{1}$, R. Bohlin$^{6}$, J. Kalirai$^{6}$}
  \\
  $^{1}$Department of Physics, University of Warwick, CV4 7AL, Coventry, UK\\
  $^{2}$Astronomisches Rechen-Institut, Zentrum f\"ur Astronomie der Universit\"at 
  Heidelberg, D-69120 Heidelberg, Germany\\
  $^{3}$Arts et M\'etiers ParisTech Centre de Bordeaux-Talence, Esplanade des
  Arts et M\'etiers, 33400 Talence, France\\
  $^{4}$Department of Physics and Astronomy, University of Sheffield, S3 7RH, Sheffield, UK\\
  $^{5}$Institut f\"ur Theoretische Physik und Astrophysik, Universit\"at Kiel, D-24098 Kiel, Germany\\
  $^{6}$Space Telescope Science Institute, 3700 San Martin Drive, Baltimore,
  MD 21218, USA }

\date{Accepted XXX. Received YYY; in original form ZZZ}

\pubyear{2016}
\begin{document}
\label{firstpage}
\pagerange{\pageref{firstpage}--\pageref{lastpage}}
\maketitle

\begin{abstract}
  The {\it Gaia} Data Release 1 (DR1) sample of white dwarf parallaxes is
  presented, including 6 directly observed degenerates and 46 white dwarfs in
  wide binaries. This data set is combined with spectroscopic atmospheric
  parameters to study the white dwarf mass-radius relationship (MRR). {\it
    Gaia} parallaxes and $G$ magnitudes are used to derive model atmosphere
  dependent white dwarf radii, which can then be compared to the predictions
  of a theoretical MRR. We find a good agreement between {\it Gaia} DR1
  parallaxes, published effective temperatures ($T_{\rm eff}$) and surface
  gravities ($\log g$), and theoretical MRRs. As it was the case for {\it
    Hipparcos}, the precision of the data does not allow for the
  characterisation of hydrogen envelope masses. The uncertainties on the
  spectroscopic atmospheric parameters are found to dominate the error budget
  and current error estimates for well-known and bright white dwarfs may be
  slightly optimistic. With the much larger {\it Gaia} DR2 white dwarf sample
  it will be possible to explore the MRR over a much wider range of mass,
  $T_{\rm eff}$, and spectral types.

\end{abstract}


\section{Introduction}

The white dwarf mass-radius relationship (MRR) is fundamental to many aspects
of astrophysics. At one end of the spectrum, the upper mass limit first
derived by \citet{chandrasekhar} is the central basis of our understanding of
type Ia supernovae, standard candles that can be used to measure the expansion
of the Universe \citep{riess98,perlmutter99}.  On the other hand, the MRR is
an essential ingredient to compute white dwarf masses from spectroscopy,
photometry, or gravitational redshift measurements \citep[see,
e.g.,][]{koester79,shipman79,koester87,bergeron92,bergeron01,falcon12}.  These
masses calibrate the semi-empirical initial to final mass relation for white
dwarfs in clusters and wide binaries \citep[see,
e.g.,][]{IFMR1,catalan08,kalirai08,williams09,casewell09,dobbie12,cummings16}.
These results unlock the potential for white dwarfs to be used to understand
the chemical evolution of galaxies \citep{kalirai14}, date old stellar
populations \citep{hansen07,kalirai12}, and trace the local star formation
history \citep{tremblay14}.

\begin{figure}
  \centering \includegraphics[scale=0.33,bb=30 80 592 722,angle=270]{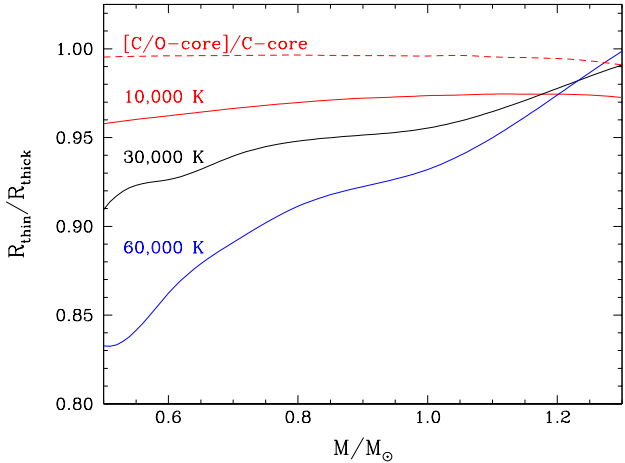}
  \caption[f1.eps]{Ratio of the predicted radii for thick ($q_{\rm H} =
    10^{-4}$) and thin ($q_{\rm H} = 10^{-10}$) hydrogen layers as a function
    of the white dwarf mass. Cooling sequences from \citet{fontaine01} at
    $T_{\rm eff}$ = 10,000 K (solid red line) and 30,000 K (black), as well as
    the models of \citet{wood95} at 60,000 K (blue) were employed. We also
    show the difference between the C/O-core (50/50 by mass fraction mixed
    uniformly) and pure-C cooling tracks at 10,000 K (dashed red line).
    \label{fg:f1}}
\end{figure}

On the theoretical side, the first MRRs that were utilized assumed a zero
temperature fully degenerate core \citep{hamada61}. The predictions have now
improved to include the finite temperature of C and O nuclei in
the interior and the non-degenerate upper layers of He and H
\citep{wood95,hansen99,fontaine01,salaris10,althaus10a}. The MRRs were also
extended to lower and higher mass ranges, with calculations for He and O/Ne
cores, respectively \citep{althaus07,althaus13}. The total mass of the
gravitationally stratified H, He, and C/O layers in white dwarfs is poorly
constrained since we can only see the top layer from the outside. While there
are some constraints on the interior structure of white dwarfs from
asteroseismology \citep{fontaine92,romero12,romero13,gia16}, the white dwarf
cooling sequence in clusters \citep{hansen15,goldsbury16}, and convective
mixing studies \citep{sion84,tremblay08,bergeron11}, a theoretical MRR
assuming a specific interior stratification is usually preferred
\citep{iben84,fontaine01,althaus10b}. For hydrogen-atmosphere DA white dwarfs,
most studies assume thick hydrogen layers with $q_{\rm H}$ = $M_{\rm H}/M_{\rm
  tot}$ = 10$^{-4}$, which is an estimate of the maximum hydrogen mass for
residual nuclear burning \citep{iben84}. More detailed calculations for the
maximum H envelope mass as a function of the white dwarf mass have also been
employed \citep{althaus10b}. On the other hand, thin H-layers ($q_{\rm H}$ =
10$^{-10}$) are often used for helium atmospheres (DB, DZ, DQ, and DC).
Fig. \ref{fg:f1} demonstrates that the MRR varies by 1--15\%, depending
  on the white dwarf mass and temperature, whether a thick or a thin hydrogen
layer is assumed. As a consequence, an observed MRR that would achieve a
1\%-level precision could in principle constrain the layering of white
dwarfs. On the other hand, Fig. \ref{fg:f1} shows that the effect of varying
the C/O ratio in the core is very small on the MRR ($< 1\%$).

Despite its fundamental importance, the MRR of white dwarfs is not robustly
constrained by observations. One of the most successful tests so far has been
from eclipsing binaries including a white dwarf. Currently, this method can
reach a precision of $\sim$2\% on the MRR \citep{parsons16}. These derivations
are based on photometric observations of the eclipses and kinematic
parameters, and are almost completely independent of white dwarf model
atmospheres. The disadvantage is that there are only a few known such systems
\citep{obrien01,parsons10,pyrzas12,parsons12a,parsons12b,parsons12c,bours15,parsons16}
and their configuration implies that they are always post-common envelope
binaries that have previously interacted.

Another method to test the MRR is to rely on astrometric binaries with known
distances and precise dynamical orbital mass measurements
\citep{shipman97,barstow05,bond15}. There are only a few such systems, with
Sirius, 40 Eri, and Procyon being the most studied. One can then use the
observed gravitational redshift, e.g. from the wavelength shift of the cores
of the Balmer lines, to derive the radius of the white dwarf relatively
independently of its atmospheric parameters. For the case of Sirius B, the
gravitational redshift measurements are still not fully understood and more
work is needed to comprehend all constraints on mass and radius
\citep{barstow05,barstow15}. Nevertheless, high-resolution and high
signal-to-noise spectroscopic observations allow for radial velocity
measurements at a $\sim$2.5\% precision level \citep{zuckerman13},
highlighting the potential of this technique.

All other methods to derive the MRR are semi-empirical and rely on the
atmospheric parameters, the effective temperature ($T_{\rm eff}$) and surface
gravity ($\log g$). The latter are most often constrained by comparing
detailed model spectra to the observed Balmer lines in DA white dwarfs
\citep{bergeron92,finley97} and to the He I lines in DB white dwarfs
\citep{bergeron11,koester15}. If a dynamical mass is available, one can then
derive the radius from the spectroscopic surface gravity, but for most white
dwarfs it is not possible.

The calculation of the semi-empirical MRR using atmospheric parameters was
pioneered by \citet{schmidt96} and \citet{vauclair97} with trigonometric
parallax measurements for 20 white dwarfs directly observed from the {\it
  Hipparcos} satellite. This technique was later expanded to include wide
binary systems for which the primary has a precise {\it Hipparcos} parallax
\citep{provencal98,holberg12}. This method is based on the fact that the
energy flux measured at the earth is $R^2/D^2$ times the flux emitted at the
surface of the star, where $R$ is the stellar radius and $D$ the distance
  to earth.  The flux emitted at the surface itself depends on the
predictions from model atmospheres. The atmospheric parameters coupled with
the distance can therefore allow for the derivation of a semi-empirical
radius. As highlighted by \citet{vauclair97}, once the surface flux is
integrated and observed over a broad photometric band, the derived radius
depends almost only on $T_{\rm eff}$ and very little on $\log g$. One can then
compute a mass independent of the MRR by using the radius defined above and
the spectroscopic $\log g$.  

Given that the atmospheric parameters are employed to derive the
semi-empirical MRR, it is not straightforward to disentangle a genuine
signature of a MRR and interior structure from systematic model atmosphere
effects. We note that some authors have actually {\it assumed} a theoretical
MRR and used the technique described above to test the accuracy of the
atmospheric parameters and model atmospheres \citep[see, e.g.,][]{tremblay13}.
To complicate matters even more, there is a partial degeneracy since
increasing both $T_{\rm eff}$ and $\log g$ can result in the same predicted
luminosity and distance \citep{tremblay09}.

Despite the fact that modern ground-based techniques have achieved a $\sim$0.5
milliarcsec (mas) precision for parallaxes of a few selected white dwarfs in
the solar neighborhood \citep{harris07,subasavage09}, the picture of the
semi-empirical MRR has remained largely unchanged since the {\it Hipparcos}
study of \citet{vauclair97} and the follow-up by \citet{holberg12}.
\citet{vauclair97} found that the {\it Hipparcos} MRR is largely consistent
with theoretical predictions when realistic uncertainties on the atmospheric
parameters are taken into account.  They concluded that the error bars on the
atmospheric parameters published in the literature at the time were slightly
too optimistic, and that the determination of the size of the H-layers for
{\it Hipparcos} white dwarfs was out of reach.

The main goal of this work is to use {\it Gaia} DR1 parallaxes for the {\it
  Hipparcos} and Tycho-2 catalog white dwarfs, both directly observed and in
wide binaries, to re-asses the semi-empirical MRR for degenerate stars. In
preparation for future {\it Gaia} data releases, we want to understand whether
it is possible to disentangle uncertainties in the spectroscopic technique
from a genuine offset between the theoretical and observed MRRs. Our study is
constructed as follows. First we introduce in Section~2 the {\it Gaia} DR1 and
{\it Hipparcos} white dwarf samples and determine the atmospheric parameters
of these objects. We derive the semi-empirical MRR in Section~3 and discuss
the implications in Section~4. We conclude in Section~5.

\section{The Gaia DR1 sample}

The European Space Agency (ESA) astrometric mission {\it Gaia} is the
successor of the {\it Hipparcos} mission and increases by orders of magnitude
the precision and number of sources. {\it Gaia} will determine positions,
parallaxes, and proper motions for $\sim$1\% of the stars in the Galaxy, and
the catalog will be complete for the full sky for V $\lesssim$ 20 mag
\citep{gaia1}. The final data release will include between 250,000 and 500,000
white dwarfs, and among those 95\% will have a parallax precision better than
10\% \citep{torres05,carrasco14}.  The final catalog will also include $G$
passband photometry, low-resolution spectrophotometry in the blue (BP,
  330--680 nm) and red (RP, 640--1000 nm), and (for bright stars, $G \lesssim 15$) higher-resolution spectroscopy
  in the region of the Ca triplet around 860 nm with the Radial Velocity
  Spectrometer \citep{jordi10,carrasco14}.

The {\it Gaia} DR1 is limited to $G$ passband photometry and the
five-parameter astrometric solution for stars in common with the {\it
  Hipparcos} and Tycho-2 catalogs \citep{dr11,dr12,dr13,dr14}. However, not
all {\it Hipparcos} and Tycho-2 stars are found in {\it Gaia} DR1 owing to
source filtering. In particular, sources with extremely blue or red colours do
not appear in the catalog \citep{dr14}. Unfortunately, this significantly 
  reduces the size of the {\it Gaia} DR1 white dwarf sample, with most of the
  bright and close single degenerates missing.

We have cross-matched the {\it Hipparcos} and Tycho-2 catalogs with Simbad as
well as the White Dwarf Catalog \citep{mccook99}. A search radius of
10$\arcsec$ around the reference coordinates was employed and all objects
classified as white dwarfs were looked at manually. Our method eliminates all
objects that are not known to be white dwarfs and wide binaries for which the
stellar remnant is at a separation larger than $\sim$10$\arcsec$ to the {\it
  Hipparcos} or Tycho-2 star.  We have identified 25 white dwarfs for which
the bright degenerate star itself is part of the {\it Hipparcos} (22 objects)
or Tycho-2 (3 objects) catalogs.  Those objects are shown in Table~\ref{fg:t1}
with $V$ magnitudes along with {\it Hipparcos} parallax values from
\citet{hipparcos} or alternative ground measurements if available in the
literature. The sample includes all {\it Hipparcos} white dwarfs studied by
\citet{vauclair97} though we have classified WD~0426+588 and WD~1544$-$377 as
wide binaries (Tables~\ref{fg:t2A} and \ref{fg:t2}) since the {\it Hipparcos}
star is actually the companion. We include WD~2117+539 for which the {\it
  Hipparcos} parallax solution was rejected during the reduction process.
WD~2007$-$303 and WD~2341+322 are {\it Hipparcos} degenerates not in
\citet{vauclair97} while WD~0439+466, WD~0621$-$376, and WD~2211$-$495 are
Tycho-2 white dwarfs. For HZ~43 (WD~1314+293), the {\it Hipparcos} parallax is
known to be inconsistent with the predicted MRR \citep{vauclair97}, and we
take instead the value from the Yale Parallax Catalog \citep{yale}. Only 6 of
the {\it Hipparcos} white dwarfs and none of the Tycho-2 degenerates are
present in {\it Gaia} DR1 owing to source filtering.  The {\it Gaia} DR1
parallaxes and $G$ magnitudes are identified in Table~\ref{fg:t1}. In addition
to the random errors available in the catalog, we have added a systematic
error of 0.3 mas \citep{dr14}.

\begin{table*}
\scriptsize
\centering
\caption{Parallaxes of Directly Observed White Dwarfs}
\setlength{\tabcolsep}{3.25pt}
\begin{tabular}{@{}ccccccccccccc}
\hline
\hline
\noalign{\smallskip}
WD & Alt. Name & HIP/Tycho ID & $\pi$ (Gaia) & $G$ (Gaia)  & $\pi$ (other)  & Ref & $V$  & Ref & SpT & $T_{\rm eff}$ & $\log (g)$ (spec) & Ref \\
 &             &  & [mas] & [mag]& [mas]  &      & [mag]&      &     & [K]         & [cm$^2$/s] & \\
 \noalign{\smallskip}
 \hline
 \noalign{\smallskip}
0046+051 & vMa 2                 & HIP 3829 & ...          & ...            & 234.60 (5.90) & 1 & 12.37 (0.02) & 4 & DZ   & 6220 (180)   & ...         & 10 \\
0148+467 & GD 279                & HIP 8709 & ...          & ...            & 64.53 (3.40)  & 1 & 12.44 (0.03) & 4 & DA   & 14,000 (280) & 8.04 (0.04) & 11 \\
0227+050 & Feige 22              & HIP 11650 & ...          & ...            & 37.52 (5.17)  & 1 & 12.78 (0.01) & 4 & DA   & 19,920 (310) & 7.93 (0.05) & 11 \\
0232+035 & Feige 24              & HIP 12031 & 13.06 (1.06) & 12.177 (0.004) & 10.90 (3.94)  & 1 & 12.41 (0.01) & 4 & DA+dM& 66,950 (1440)& 7.40 (0.07) & 11 \\
0310$-$688 & LB 3303             & HIP 14754 & ...          & ...            & 97.66 (1.85)  & 1 & 11.39 (0.01) & 5 & DA   & 16,860 (240) & 8.09 (0.04) & 11 \\
0439+466 & SH 2-216              & TYC 3343-1571-1 & ...          & ...            & 7.76 (0.33)   & 2 & 12.62 (0.03) & 6 & DAO+BP& 86,980 (2390)& 7.23 (0.08)& 11 \\
0501+527 & G 191-B2B             & HIP 23692 & ...          & ...            & 16.70 (2.97)  & 1 & 11.78 (0.01) & 4 & DA   & 60,920 (990) & 7.55 (0.05) & 11 \\
0621$-$376 & TYC 7613-1087-1     & TYC 7613-1087-1 & ...          & ...            & ...           &...& 12.09 (0.03) & 6 & DA+BP& 66,060 (1140)& 7.12 (0.05) & 11 \\
0644+375 & He 3                  & HIP 32560 & ...          & ...            & 63.53 (3.55)  & 1 & 12.06 (0.01) & 4 & DA   & 22,290 (340) & 8.10 (0.05) & 11 \\
1134+300 & GD 140                & HIP 56662 & ...          & ...            & 63.26 (3.60)  & 1 & 12.49 (0.02) & 4 & DA   & 22,470 (340) & 8.56 (0.05) & 11 \\
1142$-$645 & L 145-141           & HIP 57367 & 215.78 (0.57)& 11.410 (0.002) & 215.80 (1.25) & 1 & 11.51 (0.01) & 5 & DQ   & 7970 (220)  & ...         & 10 \\
1314+293 & HZ 43A                & HIP 64766 & 17.23 (0.77) & 12.907 (0.002) & 15.50 (3.40)  & 2 & 12.91 (0.03) & 4 & DA+dM& 56,800 (1250)& 7.89 (0.07) & 11 \\
1327$-$083 & Wolf 485A           & HIP 65877 & ...          & ...            & 57.55 (3.85)  & 1 & 12.34 (0.01) & 5 & DA   & 14,570 (240) & 7.99 (0.04) & 11 \\
1337+705 & G 238-44              & HIP 66578 & ...          & ...            & 38.29 (3.02)  & 1 & 12.77 (0.01) & 7 & DA   & 21,290 (330) & 7.93 (0.05) & 11 \\
1620$-$391 & CD $-$38 10980      & HIP 80300 & ...          & ...            & 76.00 (2.56)  & 1 & 11.01 (0.01) & 4 & DA   & 25,980 (370) & 7.96 (0.04) & 11 \\
1647+591 & G 226-29              & HIP 82257 & 91.04 (0.80) & 12.288 (0.001) & 94.04 (2.67)  & 1 & 12.24 (0.03) & 4 & DAV  & 12,510 (200) & 8.34 (0.05) & 11, 12\\
1917$-$077 & LDS 678A            & HIP 95071 & 95.10 (0.77) & 12.248 (0.001) & 91.31 (4.02)  & 1 & 12.29 (0.01) & 5 & DBQA & 10,400 (360) & ...         & 10 \\
2007$-$303 & L 565-18            & HIP 99438 & ...          & ...            & 61.09 (4.51)  & 1 & 12.24 (0.01) & 5 & DA   & 16,150 (230) & 7.98 (0.04) & 11 \\
2032+248 & Wolf 1346             & HIP 101516 & ...          & ...            & 64.32 (2.58)  & 1 & 11.55 (0.01) & 5 & DA   & 20,700 (320) & 8.02 (0.05) & 11 \\
2039$-$202 & L 711-10            & HIP 102207 & ...          & ...            & 48.22 (3.77)  & 1 & 12.40 (0.01) & 5 & DA   & 20,160 (300) & 7.98 (0.04) & 11 \\
2117+539 & G 231-40              & TYC 3953-480-1 & 57.76 (0.99) & 12.411 (0.001) & 50.70 (7.00)  & 3 & 12.33 (0.01) & 4 & DA   & 14,680 (240) & 7.91 (0.05) & 11 \\
2149+021 & G 93-48               & HIP 107968 & ...          & ...            & 37.51 (4.41)  & 1 & 12.74 (0.01) & 8 & DA   & 18,170 (270) & 8.01 (0.04) & 11 \\
2211$-$495 & TYC 8441-1261-1     & TYC 8441-1261-1 & ...          & ...            & ...           &...& 11.71 (0.01) & 9 & DA+BP& 71,530 (1530)& 7.46 (0.06) & 11 \\
2341+322   & LP 347-4            & HIP 117059 & ...          & ...            & 58.39 (11.79) & 1 & 12.93 (0.05) & 4 & DA   & 13,100 (200) & 8.02 (0.04) & 11, 12\\
 \noalign{\smallskip}
 \hline
 \noalign{\smallskip}

 \multicolumn{13}{@{}p{0.90\textwidth}@{}}{{\bf Notes.} The {\it Gaia} uncertainties include both the random errors and a systematic error of 0.3 mas \citep{dr14}. Only spectroscopic $\log g$ determinations are included and not the derivations based on the parallax measurements. DA+BP stands for a DA white dwarf with the Balmer line problem (see Section 2.1).
   {\bf References.}
   1) \citet{hipparcos},
   2) \citet{harris07},
   3) \citet{yale},
   4) \citet{vauclair97},
   5) \citet{koen10},
   6) \citet{mccook99},
   7) \citet{landolt07},
   8) \citet{landolt09},
   9) \citet{marsh97},
   10) \citet{gia12},
   11) \citet{gianninas11},
   12) \citet{tremblay13}.
\label{fg:t1}}
\end{tabular}
\end{table*}

\begin{table*}
\scriptsize
\centering
\caption{White Dwarfs in Wide Binaries: Binary Parameters}
\setlength{\tabcolsep}{3.25pt}
\begin{tabular}{@{}ccccccccccccc}
\hline
\hline
\noalign{\smallskip}
WD & Alt. Name & Primary & HIP/Tycho ID &$V$ (primary) & Sep. & Ref\\
 &             &         &              &[mag] & [arcsec]   \\
 \noalign{\smallskip}
 \hline
 \noalign{\smallskip}

0030+444   & G 172-4        & BD +43 100    & HIP 2600 & 10.28 & 28.8 & 1\\ %Holberg 2013
0042+140   & LP 466-033     & BD +13 99     & HIP 3550 & 9.79  & 62.4 & 1\\ %Holberg 2013
0148+641   & G 244-36       & G 244-37      & TYC 4040-1662-1 & 11.38 & 12.1 & 2\\ %Silvestri 2002
0220+222   & G 94-B5B       & HD 14784      & TYC 1221-1534-1 & 8.24  & 26.9 & 3\\ %Oswalt 1994
0221+399   & LP 196-060     & BD +39 539    & TYC 2835-349-1 & 9.84  & 40.5 & 1\\ %Holberg 2013
0250$-$007 & LP 591-177     & HD 17998      & TYC 4700-510-1 & 9.11  & 27.4 & 1\\ %Holberg 2013
0304+154   & LP 471-52      & LP 471-51     & TYC 1225-1388-1 & 11.49 & 20.6 & 1\\ %Holberg 2013
0315$-$011 & LP 592-80      & BD $-$01 469  & HIP 15383 & 5.37  & 46.1 & 1\\ %Holberg 2013
0355+255   & NLTT 12250     & HD 283255     & TYC 1817-1583-1  & 10.82 & 16.0 & 3\\ %Oswalt 1994
0400$-$346 & NLTT 12412     & HD 25535      & HIP 18824 & 6.73  & 64.1 & 1\\ %Holberg 2013
0413$-$077 & 40 Eri B       & 40 Eri A      & HIP 19849 & 4.43  & 83.4 & 1\\ %Holberg 2013
0415$-$594 &$\epsilon$ Ret B&$\epsilon$ Ret & HIP 19921 & 4.44  & 12.9 & 1\\ %Holberg 2013
0426+588   & Stein 2051B    & LHS 26        & HIP 21088 & 10.98 & 8.9  & 1\\ %Holberg 2013
0433+270   & G 39-27        & HD 283750     & HIP 21482 & 8.42  & 124  & 1\\ %Holberg 2013
0551+123   & NLTT 15768     & HD 39570      & HIP 27878 & 7.76  & 89.8 & 1\\ %Holberg 2013
0615$-$591 & BPM 18164      & HD 44120      & HIP 29788 & 6.43  & 40.7 & 1\\ %Holberg 2013
0642$-$166 & Sirius B       & Sirius A      & HIP 32349 & $-$1.46 & 8.1& 1\\ %Holberg 2013
0642$-$285 & LP 895-41      & CD $-$28 3358 & TYC 6533-994-1 & 10.57 & 16.1 & 1\\ %Holberg 2013
0658+712   & LP 34-137      & BD +71 380    & HIP 34082 & 9.34  & 28.7 & 1\\ %Holberg 2013
0736+053   & Procyon B      & Procyon       & HIP 37279 & 0.37  & 4.8  & 1\\ %Holberg 2013
0743$-$336 & VB 03          & HD 63077      & HIP 37853 & 5.37  & 868  & 1\\ %Holberg 2013
0751$-$252 & SCR J0753-2524 & NLTT 18618    & HIP 38594 & 9.72  & 400  & 4\\ %Zuckerman 2014
0842+490   & HD 74389B      & HD 74389      & HIP 42994 & 7.48  & 20.1 & 1\\ %Holberg 2013
0845$-$188 & LP 786-6       & NLTT 20261    & TYC 6020-1448-1 & 11.23 & 30.2 & 1\\ %Holberg 2013
1009$-$184 & WT 1759        & BD $-$17 3088 & HIP 49973 & 9.91  & 399  & 1\\ %Holberg 2013
1043$-$188 & LP 791-55      & BD $-$18 3019A& HIP 52621 & 11.21 & 7.1  & 2\\ %Silvestri 2002
1107$-$257 & LP 849-059     & HD 96941      & HIP 54530 & 8.69  & 100.2& 1\\ %Holberg 2013
1120+073   & LP 552-49      & LP 552-48     & HIP 55605 & 10.38 & 23.2 & 2\\ %Silvestri 2002
1130+189   & LP 433-6       & LP 433-7      & TYC 1438-418-2 & 11.15 & 154.5& 1\\ %Holberg 2013
1133+619   & LP 94-65       & LP 94-66      & TYC 4153-706-1 & 11.77 & 17.72& 1\\ %Holberg 2013
1209$-$060 & LP 674-029     & HD 106092     & HIP 59519 & 10.14 & 203  & 1\\ %Holberg 2013
1304+227   & SDSS J1307+2227& BD +23 2539   & TYC 1456-876-1 & 9.75  & 20.5 & 1\\ %Holberg 2013
1354+340   & G 165-B5B      & BD +34 2473   & HIP 68145 & 9.08  & 55.7 & 1\\ %Holberg 2013
1455+300   & NLTT 38926     & BD +30 2592   & HIP 73224 & 9.73  & 25.9 & 1\\ %Holberg 2013
1501+301   & LP 326-74      & LP 326-75     & TYC 2023-1076-1 & 12.14 & 88.4 & 1\\ %Holberg 2013
1542+729   & LP 42-164      & LP 42-163     & HIC 76902 & 10.85 & 18.4 & 1\\ %Holberg 2013
1544$-$377 & L 481-60       & HD 140901     & HIP 77358 & 6.01  & 14.8 & 1\\ %Holberg 2013
1554+215   & PG 1554+215    & BD +21 2850   & TYC 1502-1772-1  & 10.16 & 75.7 & 5\\ %NEW
1619+123   & PG 1619+123    & HD 147528     & HIP 80182  & 8.19  & 62.5 & 1\\ %Holberg 2013
1623+022   & NLTT 42785     & BD +02 3101   & HIP 80522 & 10.07 & 9.6  & 1\\ %Holberg 2013
1623$-$540 & L 266-196      & L 266-195     & TYC 8712-1589-1 & 11.92 & 39.7 & 2\\%Silvestri 2013
1659$-$531 & BPM 24602      & BPM 24601     & HIP 83431 & 5.29  & 113.5& 1\\ %Holberg 2013
1706+332   & G 181-B5B      & BD +33 2834   & HIP 83899 & 8.59  & 37.6 & 1\\ %Holberg 2013
1710+683   & LP 70-172      & LP 70-171     & TYC 4421-2830-1  & 11.46 & 27.8 & 1\\ %Holberg 2013
1743$-$132 & G 154-B5B      & G 154-B5A     & HIP 86938 & 11.91 & 32.2 & 2\\ %Silvestri 2002
1750+098   & G 140-B1B      & HD 162867     & TYC 1011-534-1 & 9.41  & 24.7 & 1\\ %Holberg 2013
1848+688   & NLTT 47097     & BD +68 1027   & HIP 92306  & 9.72  & 33.9 & 1\\ %Holberg 2013
2048+809   & LP 25-436      & BD +80 670    & TYC 4598-133-1  & 9.08  & 18.56& 1\\ %Holberg 2013
2054$-$050 & NLTT 50189     & Ross 193      & HIP 103393  & 11.92 & 15.5 & 6\\ %Gould 2004
2129+000   & LP 638$-$004   & BD $-$00 4234 & HIP 106335  & 9.89  & 133  & 1\\ %Holberg 2013
2154$-$512 & BPM 27606      & CD $-$51 13128& HIP 108405  & 10.49 & 28.5 & 3\\ %Oswalt 1994
PM J21117+0120 & ...        & ...  & TYC 527-72-1         & 10.65 & 33.5 & 5\\ %NEW
2217+211   & LP 460-003     & BD +20 5125   & HIP 110218  & 10.07 & 83.2 & 1\\ %Holberg 2013
HS 2229+2335 & ...          & HD 213545     & TYC 2219-1647-1  & 8.40  &110.10& 5\\ %NEW
SDSS J2245$-$1002 & PB 7181 & BD $-$10 5983 & TYC 5815-1030-1 & 10.30 & 60.4 & 5\\ %NEW
2253+054   & NLTT 55300     & GJ 4304       & HIP 113244  & 11.21 & 17.1 & 2\\ %Silvestri 2002
2253+812   & LP 002-697     & G 242-15      & TYC 4613-31-1  & 11.80 & 7.2  & 2\\ %Silvestri 2002
2253$-$081 & BD $-$08 5980B & HD 216777     & HIP 113231  & 8.01  & 41.8 & 1\\ %Holberg 2013
2258+406   & G 216-B14B     & G 216-B14A    & TYC 3220-1119-1  & 11.57 & 26.1 & 1\\ %Holberg 2013
2301+762   & LP 027-275     & HD 218028     & HIP 113786  & 8.75  & 13.4 & 1\\ %Holberg 2013
2344$-$266 & NLTT 57958     & CD $-$27 16448& HIP 117308 & 11.46 & 13.2 & 2\\ %Silvestri 2002
2350$-$083 & G 273-B1B      & BD $-$08 6206 & TYC 5831-189-1  & 11.00 & 23.7 & 1\\ %Holberg 2013

 \noalign{\smallskip}
 \hline
 \noalign{\smallskip}

 \multicolumn{12}{@{}p{0.55\textwidth}@{}}{
   {\bf References.}
   1) \citet{holberg13},
	 2) \citet{silvestri02},
	 3) \citet{oswalt94},
	 4) \citet{zuckerman14},
	 5) this work,
	 6) \citet{gould04}.
\label{fg:t2A}}
\end{tabular}
\end{table*}

\begin{table*}
\scriptsize
\centering
\caption{Parallaxes of White Dwarfs in Wide Binaries}
\setlength{\tabcolsep}{3.25pt}
\begin{tabular}{@{}cccccccccccc}
\hline
\hline
\noalign{\smallskip}
WD & Alt. Name & $\pi$ (Gaia) & $G$ (Gaia)  & $\pi$ (other)  & Ref & $V$  & Ref & SpT & $T_{\rm eff}$ & $\log (g)$ (spec) & Ref \\
 &             &  [mas] & [mag]& [mas]  &      & [mag]&      &     & [K]         & [cm$^2$/s] & \\
 \noalign{\smallskip}
 \hline
 \noalign{\smallskip}
0030+444   & G 172-4        & 13.97 (0.80) & 16.550 (0.002)& 11.22 (1.52) & 1 & 16.44 (0.05) & 2 & DA & 10,270 (150) & 8.03 (0.05) & 12, 13\\
0042+140   & LP 466-033     & 17.41 (0.57) & 18.405 (0.005)& 14.38 (1.44) & 1 & 18.79 (0.05) & 3 & DZA& 5070 (90)    & ...         & 14    \\
0148+641   & G 244-36       & 57.63 (0.70) & 13.938 (0.001)& ...          &...& 14.00 (0.05) & 2 & DA & 9000 (130)   & 8.14 (0.05) & 12, 13\\ 
0220+222   & G 94-B5B       & 12.74 (0.55) & ...           & ...          &...& 15.83 (0.05) & 2 & DA & 16,240 (280) & 8.05 (0.05) & 12    \\
0221+399   & LP 196-060     & 24.30 (0.55) & 17.071 (0.002)& ...          &...& 17.39 (0.05) & 2 & DA & 6250 (140)   & 8.30 (0.23) & 12, 13\\
0250$-$007 & LP 591-177     & 21.00 (0.89) & 16.291 (0.003)& ...          &...& 16.40 (0.05) & 2 & DA & 8410 (130)   & 8.20 (0.07) & 12, 13\\
0304+154   & LP 471-52      & ...          & 19.11 (0.01)  & ...          &...& 20.20 (0.10) & 2 & DC:& ...          & ...         & 2     \\
0315$-$011 & LP 592-80      & ...          & 17.493 (0.003)& 14.89 (0.84) & 1 & 17.20 (0.10) & 2 & DA & 7520 (260)   & 7.97 (0.45) &15, 16, 13\\
0355+255   & NLTT 12250     & 14.75 (0.57) & 18.237 (0.004)&    ...       &...& 16.80 (0.10) & 2 & DC:& ...          & ...         & 2     \\
0400$-$346 & NLTT 12412     & ...          & 17.417 (0.002)& 19.35 (0.63) & 1 & 17.82 (0.05) & 4 & DC & 5100 (100)   & ...         & 4     \\
0413$-$077 & 40 Eri B       & ...          & ...           & 200.62 (0.23)& 1 & 9.520 (0.05) & 2 & DA & 17,100 (260) & 7.95 (0.04) & 12    \\
0415$-$594 &$\epsilon$ Ret B& ...          & ...           & 54.83 (0.15) & 1 & 12.50 (0.05) & 2 & DA & 15,310 (350) & 7.88 (0.08) & 17    \\
0426+588   & Stein 2051B    & 181.50 (0.92)& ...           & 181.36 (3.67)& 1 & 12.44 (0.05) & 2 & DC & 7180 (180)   & ...         & 7     \\ 
0433+270   & G 39-27        & 57.22 (0.58) & 15.531 (0.001)& 55.66 (1.43) & 1 & 15.79 (0.06) & 5 & DA & 5630 (100)   & ...         & 7     \\
0551+123   & NLTT 15768     & ...          & 15.758 (0.002)& 8.68 (0.81)  & 1 & 15.87 (0.05) & 4 & DB & 13,200 (900) & ...         & 4     \\
0615$-$591 & BPM 18164      & ...          & ...           & 26.72 (0.29) & 1 & 14.09 (0.10) & 2 & DB & 15,750 (370) & 8.04 (0.07) & 18    \\
0642$-$166 & Sirius B       & ...          & ...           & 380.11 (1.26)& 1 & 8.440 (0.06) & 6 & DA & 25,970 (380) & 8.57 (0.04) & 12    \\
0642$-$285 & LP 895-41      & 15.34 (0.54) & 16.422 (0.002)&    ...       &...& 16.60 (0.05) & 2 & DA & 9280 (130)   & 7.87 (0.05) & 12, 13\\
0658+712   & LP 34-137      & 13.04 (0.68) & 18.627 (0.004)& 12.27 (1.37) & 1 & 19.20 (0.10) & 2 & DC & ...          & ...         & 2     \\
0736+053   & Procyon B      & ...          & ...           & 284.56 (1.26)& 1 & 10.94 (0.05) & 7 & DQZ& 7870 (430)   & ...         & 7     \\
0743$-$336 & VB 03          & ...          & ...           & 65.75 (0.51) & 1 & 16.59 (0.05) & 4 & DC & 4460 (100)   & ...         & 7     \\
0751$-$252 & SCR0753-2524   & 56.23 (0.56) & 15.99 (0.07)  & 51.52 (1.46) & 1 & 16.27 (0.05) & 7 & DA & 5090 (140)   & ...         & 7     \\
0842+490   & HD 74389B      & ...          & ...           & 8.97 (0.57)  & 1 & 15.00 (0.05) & 2 & DA & 40,250 (300) & 8.09 (0.05) & 19, 16\\
0845$-$188 & LP 786-6       & ...          & 15.648 (0.002)&    ...       &...& 15.68 (0.03) & 8 & DB & 17,470 (420) & 8.15 (0.08) & 18    \\
1009$-$184 & WT 1759        & ...          & 15.280 (0.002)& 58.20 (1.67) & 1 & 15.44 (0.05) & 7 & DZ & 6040 (360)   & ...         & 7     \\
1043$-$188 & LP 791-55      & 52.59 (0.69) & ...           & 49.95 (2.26) &...& 15.52 (0.05) & 7 & DQpec & 5780 (90) & ...         & 7     \\
1107$-$257 & LP 849-059     & 24.18 (0.55) & 17.273 (0.002)& 24.90 (0.98) & 1 & 16.79 (0.05) & 2 & DC & ...          & ...         & 2     \\
1120+073   & LP 552-49      & ...          & 17.159 (0.003)& 31.12 (2.35) & 1 & 17.49 (0.05) & 2 & DC & 4460 (110)   & ...         & 20    \\
1130+189   & LP 433-6       & 4.63 (0.73)  & 17.569 (0.003)&    ...       &...& 17.60 (0.10) & 2 & DA & 10,950 (190) & 8.34 (0.06) & 12, 13\\
1133+619   & LP 94-65       & 7.05 (0.80)  & 18.358 (0.002)&    ...       &...& 17.70 (0.10) & 2 & DZ & ...          & ...         & 2     \\
1209$-$060 & LP 674-029     & 22.69 (0.79) & 16.878 (0.004)& 22.18 (1.49) & 1 & 17.26 (0.05) & 2 & DA & 6590 (100)   & 8.02 (0.22) & 4     \\
1304+227   & SDSS J1307+2227& 12.96 (0.58) & 16.491 (0.002)&    ...       &...& 16.20 (0.10) & 2 & DA & 10,280 (180) & 8.21 (0.09) & 12, 13\\
1354+340   & G 165-B5B      & 10.79 (0.58) & 16.023 (0.004)& 10.06 (1.15) & 1 & 16.17 (0.01) & 5 & DA & 14,490 (290) & 8.06 (0.05) & 12    \\
1455+300   & NLTT 38926     & 15.48 (0.55) & 18.418 (0.004)& 16.51 (1.66) & 1 & 20.16 (0.10) & 9 & ...& ...          & ...         & 9     \\
1501+301   & LP 326-74      & 12.56 (1.06) & 17.654 (0.001)&    ...       &...& 17.70 (0.10) & 2 & DC & 7250         & ...         & 21    \\
1542+729   & LP 42-164      & 13.44 (0.52) & 18.077 (0.004)& 16.10 (2.48) & 1 & 18.06 (0.05) & 10& DC & ...          & ...         & 10    \\
1544$-$377 & L 481-60       & 65.57 (0.74) & 13.003 (0.001)& 65.13 (0.40) & 1 & 12.80 (0.05) & 2 & DA & 10,380 (150) & 7.96 (0.04) & 12, 13\\
1554+215   & PG 1554+215    & 9.73 (0.68)  & ...           &    ...       &...& 15.26 (0.01) & 5 & DA & 27,320 (410) & 7.90 (0.05) & 12    \\
1619+123   & PG 1619+123    & 17.70 (0.53) & ...           & 19.29 (1.02) & 1 & 14.66 (0.05) & 2 & DA & 17,150 (260) & 7.87 (0.04) & 12    \\
1623+022   & NLTT 42785     & 20.59 (0.61) & 17.50 (0.01)  & 17.64 (2.12) & 1 & 17.42 (0.05) & 9 & DC & ...          & ...         & 10    \\
1623$-$540 & L 266-196      & 21.82 (0.66) & 15.445 (0.002)& ...          &...& 15.74 (0.05) & 2 & DA & 11,280 (170) & 7.95 (0.04) & 12, 13\\
1659$-$531 & BPM 24602      & ...          & ...           & 36.73 (0.63) & 1 & 13.47 (0.05) & 2 & DA & 15,570 (230) & 8.07 (0.04) & 12    \\
1706+332   & G 181-B5B      & 13.98 (0.53) & 15.970 (0.002)& 14.35 (0.87) & 1 & 15.90 (0.05) & 2 & DA & 13,560 (390) & 7.94 (0.06) & 12, 13\\
1710+683   & LP 70-172      & 17.98 (0.75) & 17.259 (0.007)&    ...       &...& 17.50 (0.05) & 2 & DA & 6630 (230)   & 7.86 (0.51) & 12, 13\\
1743$-$132 & G 154-B5B      & 25.97 (0.75) & 14.604 (0.002)& 29.96 (3.63) & 1 & 14.22 (0.05) & 2 & DA & 12,920 (210) & 8.01 (0.05) & 12, 13\\
1750+098   & G 140-B1B      & 22.80 (0.53) & 15.615 (0.002)&    ...       &...& 15.72 (0.05) & 2 & DA & 9520         & ...         & 12    \\
1848+688   & NLTT 47097     & 11.09 (0.52) & 17.342 (0.004)& 12.68 (0.76) & 1 & 17.18 (0.05) & 9 & ...& ...          & ...         & 9     \\
2048+809   & LP 25-436      & 11.67 (1.02) & 16.434 (0.002)&    ...       &...& 16.59 (0.05) & 2 & DA & 8450 (130)   & 8.11 (0.07) & 12, 13\\
2054$-$050 & NLTT 50189     & 62.15 (0.73) & ...           & 56.54 (3.92) & 1 & 16.69 (0.05) & 7 & DC & 4340 (80)    & ...         & 7     \\
2129+000   & LP 638-004     & 23.16 (0.52) & ...           & 22.13 (2.01) & 1 & 14.67 (0.03) & 8 & DB & 14,380 (350) & 8.26 (0.14) & 18    \\
2154$-$512 & BPM 27606      & 66.13 (0.75) & 14.477 (0.001)& 62.61 (2.92) & 1 & 14.74 (0.03) & 7 & DQP& 7190 (90)    & ...         & 7     \\
PM J21117+0120 & ...        & 16.37 (1.00) & 15.266 (0.002)& ...          &...& ...          &...& DA & 16,570 (100) & 8.06 (0.05) & 20    \\
2217+211   & LP 460-003     & 18.76 (0.60) & 17.672 (0.004)& 20.30 (1.40) & 1 & 17.69 (0.05) & 2 & DC & ...          & ...         & 22    \\
HS 2229+2335 & ...          & 9.02 (0.85)  & 15.992 (0.004)&     ...      &...& 16.01 (0.09) & 5 & DA & 20,000 (500) & 7.96 (0.09) & 23, 16\\
SDSS J2245$-$1002 & PB 7181 & 16.72 (1.29) & ...           &   ...        &...& 17.02 (0.05) & 11& DA & 8700 (30)    & 8.36 (0.04) & 11, 13\\
2253+054   & NLTT 55300     & 40.06 (1.09) & ...           & 40.89 (2.12) & 1 & 15.71 (0.05) & 2 & DA & 6240 (150)   & 8.60 (0.24) & 12, 13\\
2253+812   & LP 002-697     & ...          & 17.543 (0.003)&    ...       &...& 17.30 (0.10) & 2 & DC:& ...          & ...         & 2     \\
2253$-$081 & BD $-$08 5980B & 27.97 (0.54) & 16.311 (0.002)& 27.22 (1.12) & 1 & 16.50 (0.05) & 2 & DA & 6770 (130)   & 7.82 (0.18) & 12, 13\\
2258+406   & G 216-B14B     & 13.96 (0.73) & 16.676 (0.002)&    ...       &...& 15.50 (0.10) & 2 & DA & 9910 (150)   & 8.16 (0.06) & 12, 13\\
2301+762   & LP 027-275     & 15.60 (0.56) & ...           & 14.97 (0.79) & 1 & 16.35 (0.05) & 2 & DC & ...          & ...         & 24    \\
2344$-$266 & NLTT 57958     & 21.50 (0.55) & 16.673 (0.008)& 20.03 (3.04) & 1 & 16.59 (0.05) & 2 & DB:& ...          & ...         & 2     \\
2350$-$083 & G 273-B1B      & 9.96 (1.13)  & ...           &   ...        &...& 16.18 (0.10) & 2 & DA & 19,270 (310) & 7.90 (0.05) & 12    \\
 \noalign{\smallskip}
 \hline
 \noalign{\smallskip}

 \multicolumn{12}{@{}p{0.85\textwidth}@{}}{{\bf Notes.} The {\it Gaia} uncertainties include both the random errors and a systematic error of 0.3 mas \citep{dr14}. Only spectroscopic $\log g$ determinations are included and not the derivations based on the parallax measurements. Spectral types with the ":" symbol are uncertain. 
   {\bf References.}
   1) \citet{hipparcos},
   2) \citet{mccook99},
   3) \citet{kilic10},
   4) \citet{kawka10},
   5) \citet{usno},
   6) \citet{holberg84},
   7) \citet{gia12},
   8) \citet{landolt07},
   9) \citet{gould04},
   10) \citet{holberg13},
   11) \citet{tremblay11},
   12) \citet{gianninas11},
   13) \citet{tremblay13},
   14) \citet{kilic10},
   15) \citet{catalan08},
   16) \citet{tremblay09},
   17) \citet{farihi11},
   18) \citet{bergeron11},
   19) \citet{vennes97},
   20) \citet{limoges15},
   21) \citet{girven11},
   22) \citet{hintzen86},
   23) \citet{koester09},
   24) \citet{greenstein84}.
\label{fg:t2}}
\end{tabular}
\end{table*}

Our limited search radius of 10$\arcsec$ around {\it Hipparcos} and Tycho-2
coordinates, which was designed to recover all white dwarfs that are directly
in {\it Gaia} DR1, does not allow to build a meaningful sample of wide
binaries. A list of white dwarfs that are in common proper motion pairs with
{\it Hipparcos} or Tycho-2 stars was compiled from the literature
\citep{silvestri02,gould04,holberg13,zuckerman14}. Our aim is not to have a
complete sample but rather to include most known {\it Gaia} DR1 stars with
wide degenerate companions. The 62 selected binary systems are identified in
Table~\ref{fg:t2A} along with their angular separation. Among those, 39 are
primary stars with {\it Hipparcos} parallaxes collected in Table~\ref{fg:t2},
and 23 are Tycho-2 stars with no prior distance measurements. We have found 46
of these primary stars in {\it Gaia} DR1, with parallaxes identified in
Table~\ref{fg:t2}. The resulting physical separations lead to orbital
  periods longer than those of Procyon and Sirius ($> 40$ yr), hence these
  orbital motions should have a minor impact on parallax determinations. We
can derive the semi-empirical MRR for members of wide binaries in the same way
as we do for directly observed white dwarfs. {\it Gaia} DR1 $G$ magnitudes are
available for 43 of the white dwarf companions, while $V$ magnitudes can be
found in the literature for most systems.

Our search has also recovered a large number of white dwarfs in unresolved
binaries, often in Sirius-like systems where the degenerate star is only
visible in the UV \citep{holberg03}. Whenever there was no optical
spectroscopy for these objects, we have neglected them from our sample, since
their atmospheric parameters are significantly less precise than for the white
dwarfs identified in Tables~\ref{fg:t1} and \ref{fg:t2}. This includes
WD~1736+133 and WD~1132$-$325, even though they are separated by more than
4$\arcsec$ from their bright companion \citep{holberg13}.

\subsection{Spectroscopic Parameters}

Precise atmospheric parameters determined from spectroscopic fits are a
critical ingredient to extract the semi-empirical MRR.  As a consequence, we
have ensured that we have a homogeneous determination of the atmospheric
parameters by using the same models and fitting technique for the whole sample
as much as feasible. Whenever possible, atmospheric parameters for DA white
dwarfs are taken from \citet{gianninas11}, or in a few cases from
\citet{tremblay11} and \citet{limoges15}. These studies are based on the model
spectra from \citet{tremblay11}, and 3D corrections from \citet{tremblay13}
were applied when appropriate. The uncertainties in \citet{gianninas11} are
the sum of the formal $\chi^2$ errors and external errors of 1.2$\%$ in
$T_{\rm eff}$ and 0.038 dex in $\log g$.  The latter were determined by
observing selected stars on different nights and at different sites
\citep{LBH}. There are five DA white dwarfs, all in wide binaries, that are
not part of the \citet{gianninas11} sample. For WD~0315$-$011, $\epsilon$ Ret
B, WD~0842+490, WD 1209$-$060, and HS 2229+2335, we use the atmospheric
parameters of \citet{catalan08}, \citet{farihi11}, \citet{vennes97},
\citet{kawka10}, and \citet{koester09}, respectively. Except for
  \citet{farihi11}, these studies were performed prior to the inclusion of the
  \citet{tremblay09} Stark profiles, hence we have corrected for this effect
  using fig. 12 of \citet{tremblay09} and added 3D corrections when
  appropriate. Finally, WD~0221+399, WD~0433+270, WD~751$-$252, WD~1750+098,
and WD~2253+054 have very weak Balmer lines, hence they have no spectroscopic
gravities.

A few hot white dwarfs that are identified with spectral type DA+BP (or
DAO+BP) have the so-called Balmer line problem \citep{werner96}. In those
cases, the \citet{gianninas11} solution is with CNO added to the model
atmospheres. We also note that the optical spectrum of HZ~43 employed by
\citet{gianninas11} shows some evidence of contamination from the close M
dwarf companion. As a consequence, the error bars for this star should be
taken with some caution.

For the DB white dwarfs WD~0615$-$591, WD~0845$-$188, and WD~2129+004, we use
the atmospheric parameters from \citet{bergeron11}. Even though they are in
the regime $T_{\rm eff} < 16,000$ K, where spectroscopic $\log g$
determinations are unreliable \citep{bergeron11, koester15}, we keep them in
the sample as Section~3 demonstrates that they are in agreement with the
theoretical MRRs when parallaxes are available. However, we make no attempt to
determine whether a thin H-layer is more appropriate for these objects, as
suggested from the lack of hydrogen at the surface. On the other hand,
WD~0551+123 and WD~1917$-$077 are too cool for a meaningful $\log g$
determination from the He I lines.

For 15 DC, 1 probable DB, 4 DQ, 4 DZ, and 2 probable white dwarfs, there are
no spectroscopic $\log g$ determinations, hence no independent mass
determinations apart from using the parallaxes and magnitudes from
Tables~\ref{fg:t1} and \ref{fg:t2}, combined with a theoretical MRR. We do not
perform such mass determinations as it is outside the scope of this work to
review the photometric fits of these objects. We only include the 48 DA and 2
DB white dwarfs with spectroscopic $\log g$ values and at least one parallax
measurement in our analysis.

\section{The Mass-Radius Relation}

We employ the method of \citet{vauclair97} to study the semi-empirical
  MRR. The first step is to define the surface flux in erg sec$^{-1}$
  cm$^{-2}$ \AA$^{-1}$ from the predicted emergent monochromatic Eddington
  flux $H_{\lambda}$,  

\begin{equation}
F_{\rm surface} = 4\pi H_{\lambda}(T_{\rm eff},\log g)~,
		\label{eq1}
\end{equation}

{\noindent}where we have explicitly included the dependence on the atmospheric
parameters. The flux measured at the earth is

\begin{equation}
  f_{\rm earth} = \frac{R^2}{D^2}F_{\rm surface}~,
		\label{eq1A}
\end{equation}

{\noindent}which fully accounts for limb-darkening. However, the flux is
usually integrated over some characteristic photometric passband, such as
Johnson-Kron-Cousins $V$ or {\it Gaia} $G$, and measured by a photon-counting
device. Conversely, a surface magnitude $m_{\rm o}$ can be predicted

\begin{equation}
  m_{\rm o} = -2.5\log\left( \frac{\int S(\lambda)F_{\rm surface}\lambda d\lambda}{\int S(\lambda)\lambda d\lambda} \right)+C_S~,
  \label{eq1B}
\end{equation}

{\noindent}where $S(\lambda)$ is the total system quantum efficiency and $C_s$
is the zero point. The zero point for the $V$ filter is defined from the Vega
magnitude of +0.026 which results in $C_{V} = -$21.0607 \citep{holberg06}. If
we use the same procedure as \citet{holberg06} for the {\it Gaia} $G$ filter
where Vega has a magnitude of +0.03 \citep{jordi10}, we obtain $C_{G} =
-$21.48050. The radius is then found from

\begin{equation}
  \log R/R_{\odot} = 0.2 (m_{\rm o}-m) - \log \pi[{\rm arcsec}] + 7.64697~,
  \label{eq2}
\end{equation}

{\noindent}where $\pi$ is the trigonometric parallax in arcsec, $m$ is the
apparent magnitude, and the constant is $\log({\rm parsec}/R_{\odot})$.

A correction for interstellar extinction could be necessary for white dwarfs
with parallaxes smaller than about 20 mas \citep{genest14}. For the
magnitude-limited directly observed {\it Hipparcos} white dwarf sample, this
corresponds to $T_{\rm eff} \gtrsim 50,000$~K, including G191$-$B2B which is
suggested to have a small reddening of $E(B-V) =$ 0.0005
\citep{bohlin14}. Nevertheless, it is difficult to calculate individual
corrections that would be appropriate for our sample, and we neglect this
effect.

The emergent fluxes from the model atmospheres of \citet{tremblay11} were
integrated over the {\it Gaia} $G$ passband using Eq.~\ref{eq1B} as was done
in the preparatory work of \citet{carrasco14}. The resulting radii $R_{\rm
  Gaia}$ from Eq.~\ref{eq2} are given in Table~\ref{fg:t3}. The results using
instead the {\it Hipparcos} or ground-based parallaxes ($R_{\rm Hipparcos}$)
are also shown in Table~\ref{fg:t3}.  In those cases, we have still employed
the apparent {\it Gaia} $G$ magnitudes when available.

\begin{figure}
  \centering \includegraphics[scale=0.33,bb=30 80 492 722,angle=270]{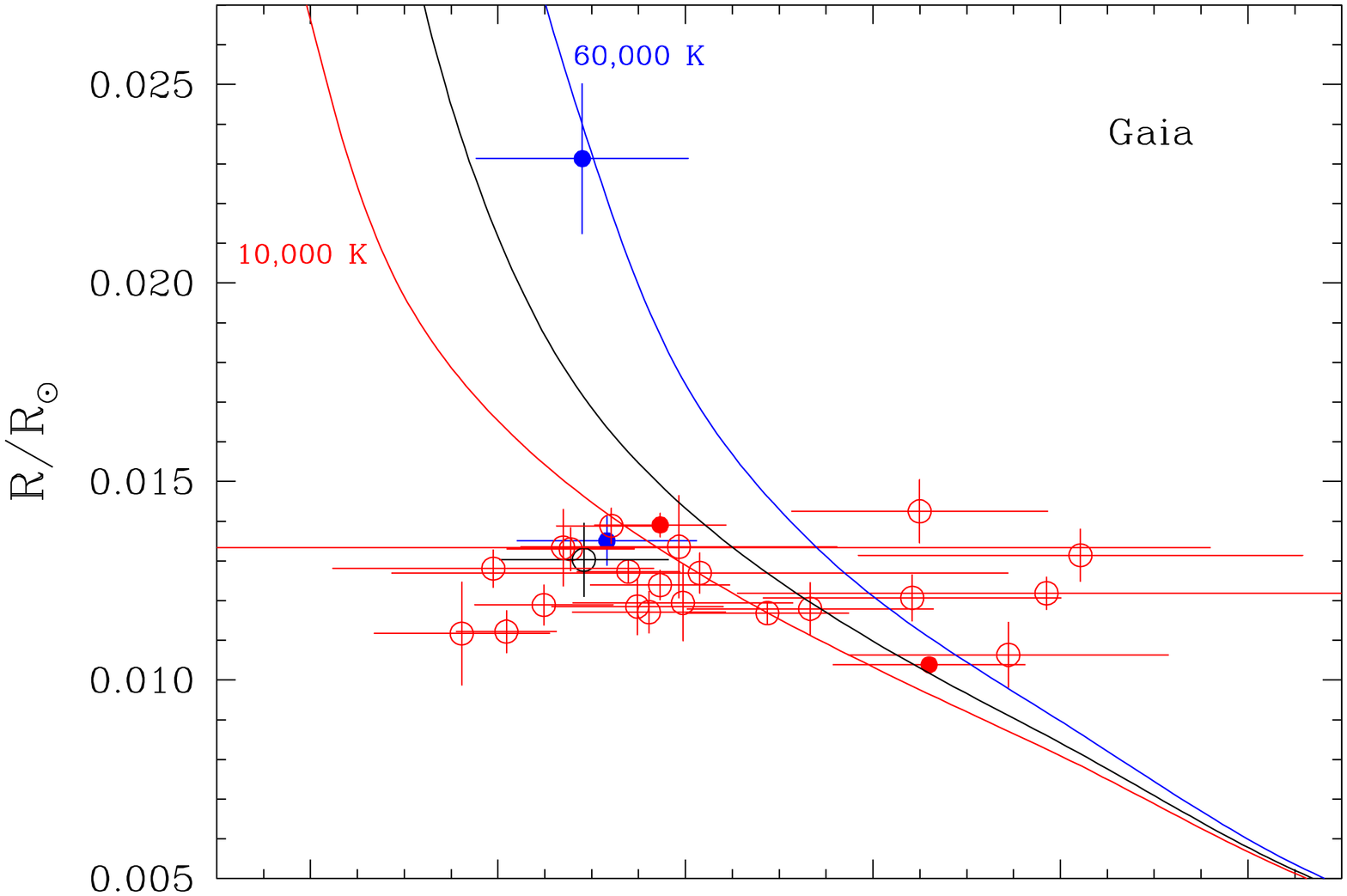} 
  \centering \includegraphics[scale=0.33,bb=30 80 592 722,angle=270]{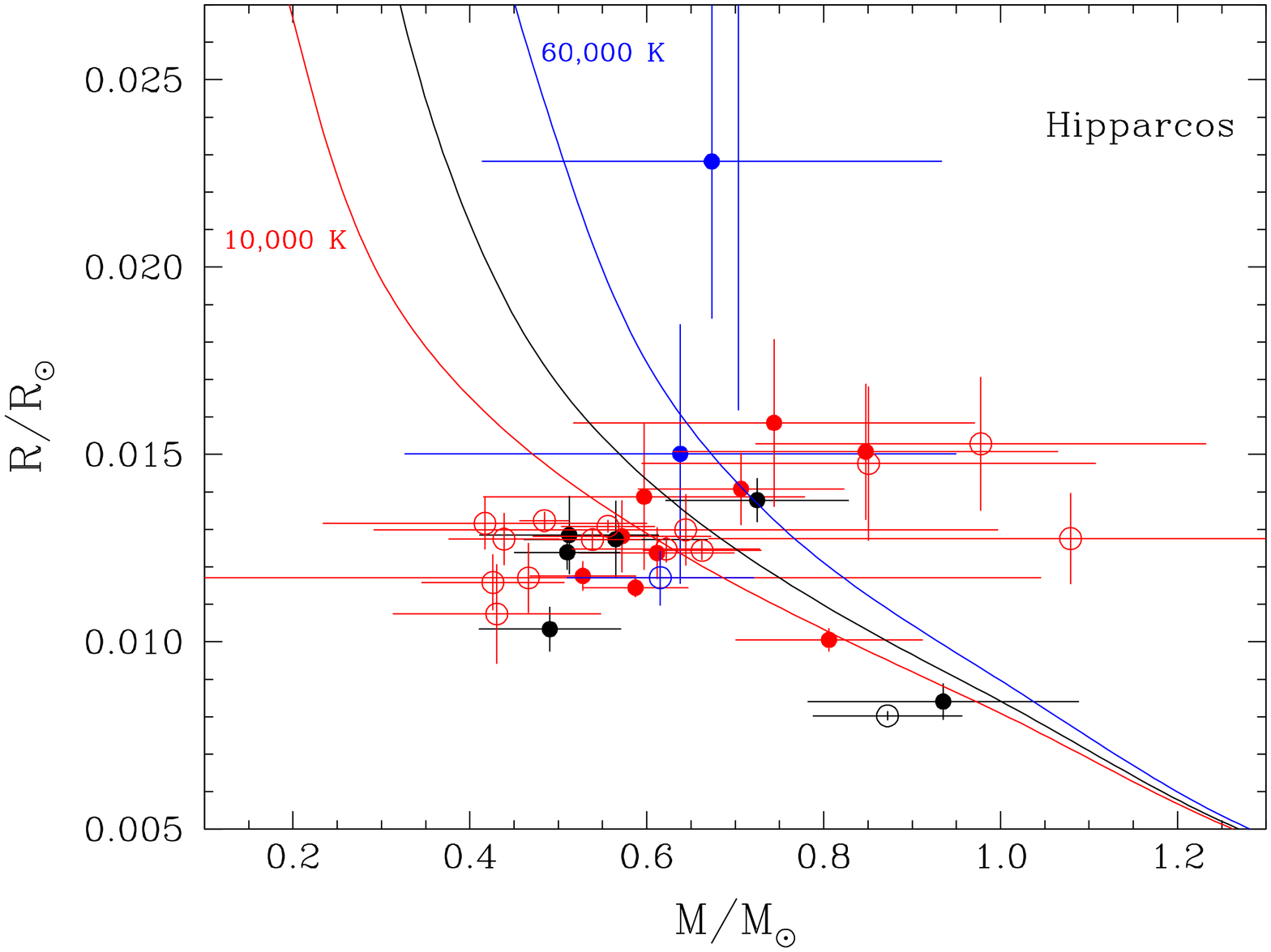}
  \caption[f2.eps]{{\it (Top:)} semi-empirical MRR using {\it Gaia} DR1
    and atmospheric parameters defined in Table~\ref{fg:t1} for directly
    observed white dwarfs (solid circles) and in Table~\ref{fg:t2} for wide
    binaries (open circles). Numerical values are given in
    Table~\ref{fg:t3}. Theoretical MRRs for $q_{\rm H} = 10^{-4}$
    \citep{wood95,fontaine01} at 10,000~K (red), 30,000~K (black), and
    60,000~K (blue) are also shown. The data points are also colour coded
    based on their $T_{\rm eff}$ and the closest corresponding theoretical
    sequence.  {\it (Bottom:)} Similar to the top panel but with pre-{\it
      Gaia} parallax measurements (mostly from {\it Hipparcos}) identified in
    Tables~\ref{fg:t1} and \ref{fg:t2}. We still rely on {\it Gaia} $G$
    magnitudes when available.
    \label{fg:f2}}
\end{figure}

Traditionally, the next step has been to compute a mass independently of the
MRR by combining the radii determined above with the spectroscopic $\log
g$. These masses are given in Table~\ref{fg:t3} and presented in a M-R diagram
in Fig.~\ref{fg:f2} for both the {\it Gaia} DR1 (top panel) and {\it
  Hipparcos} parallaxes (bottom panel). We note that the errors typically form
elongated ellipses \citep{holberg12} corresponding to the fact that $M$ is a
function of $R^{2}$. Furthermore, the predicted positions on the M-R diagram
depend on $T_{\rm eff}$, as illustrated in Fig.~\ref{fg:f2} by the theoretical
MRRs from \citet{wood95} and \citet{fontaine01} with thick H-layers at 10,000,
30,000, and 60,000~K. For these reasons, it is not straightforward to
interpret the results in a M-R diagram. In particular, the data points in
Fig.~\ref{fg:f2}, both for the {\it Gaia} DR1 and {\it Hipparcos} samples, do
not form a clear sequence of decreasing radius as a function of increasing
mass as in the predicted MRR.  This is in part caused by observational uncertainties, 
the fact that most white dwarfs in the sample have similar masses around $\sim$0.6 
$M_{\odot}$, and that for a given mass the radius will change as a function of $T_{\rm eff}$.

WD~1130+189 and WD~2048+809 are two peculiar white dwarfs in {\it Gaia} DR1
for which the observed radii $R_{\rm Gaia}$ are about twice the predicted
values. Given the surface gravities, this would lead to spurious observed masses well above the
Chandrasekhar mass limit. The natural explanation for this behaviour is that
these wide binaries are actually rare triple systems with unresolved double
degenerates \citep{obrien01,triple2,triple3}. These white dwarfs had no
parallax measurements until now and were not known to be double
degenerates. However, high-resolution observations of WD~2048+809 show
peculiar line cores that can not be explained by rotation or magnetic fields
\citep{spy05}. \citet{liebert91} and \citet{tremblay11} have shown that double
DA white dwarfs can almost perfectly mimic a single DA in spectroscopic and
photometric analyses. As a consequence, it may not be surprising that {\it
  Gaia} is able to reveal for the first time the double degenerate nature of
these objects.

\begin{table*}
\scriptsize
\centering
\caption{Semi-Empirical White Dwarf Mass-Radius Relation}
\setlength{\tabcolsep}{3.25pt}
\begin{tabular}{@{}cccccc}
\hline
\hline
\noalign{\smallskip}
WD & $M_{\rm Hipparcos}$ & $R_{\rm Hipparcos}$ & $M_{\rm Gaia}$ & $R_{\rm Gaia}$ & $R_{\rm MRR}$ \\
   & [$M_{\odot}$]     & [$0.01R_{\odot}$]  &  [$M_{\odot}$]  & [$0.01R_{\odot}$]  & [$0.01R_{\odot}$] \\

   \noalign{\smallskip}
   \hline
   \noalign{\smallskip}
   \multicolumn{5}{l}{Directly~Observed~White~Dwarfs} \\
   \hline
   \noalign{\smallskip}
   0148+467   & 0.612 (0.088) &   1.237 (0.068) & ...           & ...           & 1.259 (0.034) \\
   0227+050   & 0.597 (0.182) &   1.387 (0.196) & ...           & ...           & 1.372 (0.048) \\
   0232+035   & 0.703 (0.596) &   2.771 (1.153) & 0.490 (0.113) & 2.313 (0.190) & 2.384 (0.151) \\
   0310$-$688 & 0.587 (0.060) &   1.144 (0.025) & ...           & ...           & 1.221 (0.034) \\
   0439+466   & 0.506 (0.104) &   2.858 (0.126) & ...           & ...           & 2.960 (0.226) \\
   0501+527   & 0.674 (0.260) &   2.282 (0.419) & ...           & ...           & 2.049 (0.091) \\
   0644+375   & 0.490 (0.080) &   1.034 (0.059) & ...           & ...           & 1.222 (0.044) \\
   1134+300   & 0.935 (0.153) &   0.840 (0.049) & ...           & ...           & 0.857 (0.036) \\
   1314+293   & 0.638 (0.312) &   1.501 (0.346) & 0.516 (0.096) & 1.351 (0.062) & 1.522 (0.087) \\
   1327$-$083 & 0.706 (0.117) &   1.408 (0.096) & ...           & ...           & 1.304 (0.035) \\
   1337+705   & 0.512 (0.101) &   1.285 (0.103) & ...           & ...           & 1.376 (0.048) \\
   1620$-$391 & 0.510 (0.060) &   1.239 (0.045) & ...           & ...           & 1.360 (0.039) \\
   1647+591   & 0.806 (0.106) &   1.005 (0.031) & 0.860 (0.103) & 1.038 (0.016) & 1.013 (0.038) \\
   2007$-$303 & 0.572 (0.101) &   1.282 (0.096) & ...           & ...           & 1.316 (0.036) \\
   2032+248   & 0.725 (0.104) &   1.378 (0.059) & ...           & ...           & 1.291 (0.046) \\
   2039$-$202 & 0.565 (0.104) &   1.274 (0.102) & ...           & ...           & 1.326 (0.037) \\
   2117+539   & 0.744 (0.227) &   1.584 (0.224) & 0.573 (0.071) & 1.391 (0.030) & 1.376 (0.046) \\
   2149+021   & 0.847 (0.218) &   1.507 (0.181) & ...           & ...           & 1.294 (0.036) \\
   2341+322   & 0.528 (0.060) &   1.176 (0.039) & ...           & ...           & 1.274 (0.035) \\
   \noalign{\smallskip}
   \hline
   \noalign{\smallskip}
   \multicolumn{5}{l}{White~Dwarfs~in~Wide~Binaries}\\
   \hline
   \noalign{\smallskip}
   0030+444   & 0.851 (0.257) &   1.476 (0.206) & 0.549 (0.092) & 1.185 (0.072) & 1.258 (0.042) \\
   0148+641   & ...           &   ...           & 0.687 (0.087) & 1.169 (0.030) & 1.165 (0.040) \\
   0220+222   & ...           &   ...           & 0.561 (0.082) & 1.171 (0.053) & 1.255 (0.044) \\
   0250$-$007 & ...           &   ...           & 0.842 (0.159) & 1.207 (0.059) & 1.116 (0.055) \\
   0315$-$011 & 0.466 (0.579) &   1.171 (0.093) & ...           & ...           & 1.300 (0.382) \\
   0413$-$077 & 0.556 (0.053) &   1.308 (0.016) & ...           & ...           & 1.346 (0.037) \\
   0415$-$594 & 0.484 (0.028) &   1.323 (0.024) & ...           & ...           & 1.406 (0.019) \\
   0615$-$591 & 0.622 (0.106) &   1.247 (0.035) & ...           & ...           & 1.263 (0.061) \\
   0642$-$166 & 0.872 (0.084) &   0.802 (0.012) & ...           & ...           & 0.851 (0.029) \\
   0642$-$285 & ...           &   ...           & 0.478 (0.068) & 1.329 (0.055) & 1.392 (0.044) \\
   0842+490   & 0.615 (0.106) &   1.171 (0.075) & ...           & ...           & 1.266 (0.049) \\
   1130+189   & ...           &   ...           & ...           & 2.061 (0.336) & 1.012 (0.046) \\
   1209$-$060 & 0.644 (0.353) &   1.299 (0.094) & 0.615 (0.329) & 1.270 (0.051) & 1.256 (0.180) \\
   1304+227   & ...           &   ...           & 1.021 (0.237) & 1.314 (0.067) & 1.111 (0.071) \\
   1354+340   & 0.978 (0.255) &   1.528 (0.179) & 0.850 (0.137) & 1.425 (0.080) & 1.243 (0.043) \\
   1544$-$377 & 0.539 (0.055) &   1.273 (0.029) & 0.539 (0.055) & 1.273 (0.029) & 1.318 (0.035) \\
   1554+215   & ...           &   ...           & 0.492 (0.090) & 1.303 (0.093) & 1.424 (0.052) \\
   1619+123   & 0.439 (0.063) &   1.274 (0.069) & 0.521 (0.059) & 1.388 (0.045) & 1.421 (0.039) \\
   1623$-$540 & ...           &   ...           & 0.409 (0.054) & 1.122 (0.053) & 1.330 (0.035) \\
	 1659$-$531 & 0.663 (0.067) &   1.244 (0.026) & ...           & ...           & 1.236 (0.034) \\
   1706+332   & 0.426 (0.081) &   1.158 (0.075) & 0.449 (0.074) & 1.189 (0.052) & 1.345 (0.054) \\
   1710+683   & ...           &   ...           & 0.470 (0.691) & 1.333 (0.097) & 1.390 (0.449) \\
   1743$-$132 & 0.430 (0.117) &   1.074 (0.133) & 0.573 (0.075) & 1.239 (0.038) & 1.282 (0.043) \\
   2048+809   & ...           &   ...           & ...           & 2.018 (0.184) & 1.188 (0.057) \\
	PM J21117+0120& ...         &   ...           & 0.597 (0.118) & 1.194 (0.096) & 1.247 (0.044) \\
   2129+000   & 1.079 (0.409) &   1.275 (0.121) & 0.985 (0.330) & 1.219 (0.042) & 1.078 (0.111) \\
   HS 2229+2335& ...          &   ...           & 0.593 (0.169) & 1.336 (0.130) & 1.344 (0.085) \\
   SDSS J2245$-$1002 & ...    &   ...           & 0.944 (0.171) & 1.063 (0.083) & 0.994 (0.030) \\
   2253$-$081 & 0.417 (0.183) &   1.316 (0.069) & 0.395 (0.171) & 1.281 (0.049) & 1.425 (0.158) \\
   2258+406   & ...           &   ...           & 0.733 (0.132) & 1.179 (0.067) & 1.151 (0.048) \\
   2350$-$083 & ...           &   ...           & 0.361 (0.093) & 1.117 (0.129) & 1.399 (0.048) \\
   \noalign{\smallskip}
   \hline
   \noalign{\smallskip}
\label{fg:t3}
\end{tabular}
\end{table*}

In the following, we compare the {\it observed} radius $R_{\rm Gaia}$ or
$R_{\rm Hipparcos}$ defined by Eq.~\ref{eq2} to a {\it predicted} radius
$R_{\rm MRR}$ drawn from theoretical MRRs and spectroscopic atmospheric
parameters, an approach also favoured by \citet{holberg12}. We note that
neither quantity is purely observed or purely predicted and both depend on
the spectroscopic atmospheric parameters, hence model
atmospheres. Nevertheless, $R_{\rm Gaia}$ depends almost only on $T_{\rm eff}$
while $R_{\rm MRR}$ depends largely on $\log g$. Theoretical MRRs with thick
H-layers ($q_{\rm H} = 10^{-4}$) were employed for our standard
derivation. For $M > 0.45$ $M_{\rm \odot}$, we use the evolutionary sequences
of \citet[][$T_{\rm eff} \leq$ 30,000 K, C/O-core 50/50 by mass fraction mixed
uniformly]{fontaine01} and \citet[][$T_{\rm eff} >$ 30,000 K, pure
C-core]{wood95}. For lower masses we use the He-core sequences of
\citet{althaus01}.

\begin{figure}
  \centering \includegraphics[scale=0.33,bb=30 80 512
  722,angle=270]{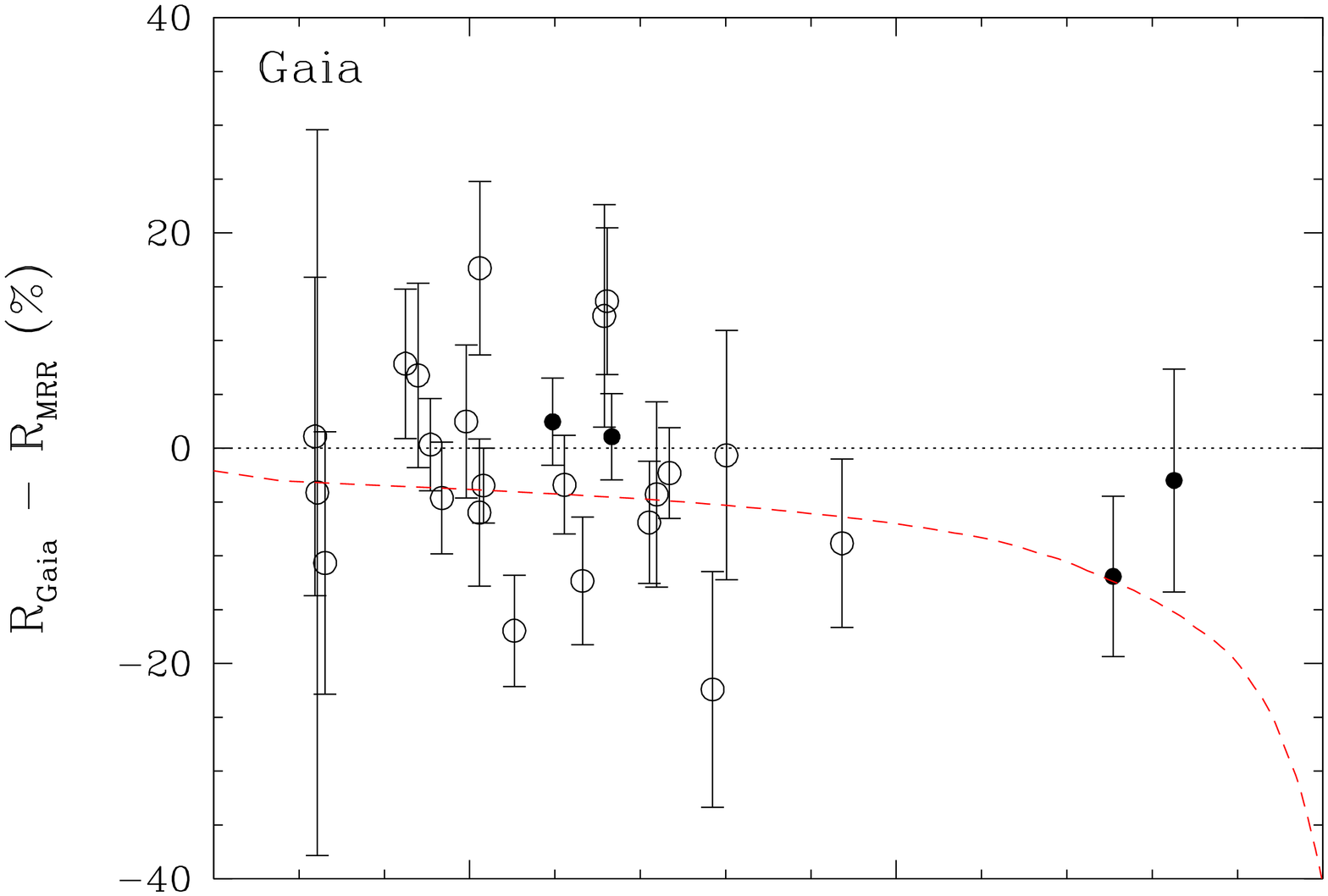} \centering \includegraphics[scale=0.33,bb=30 80
  592 722,angle=270]{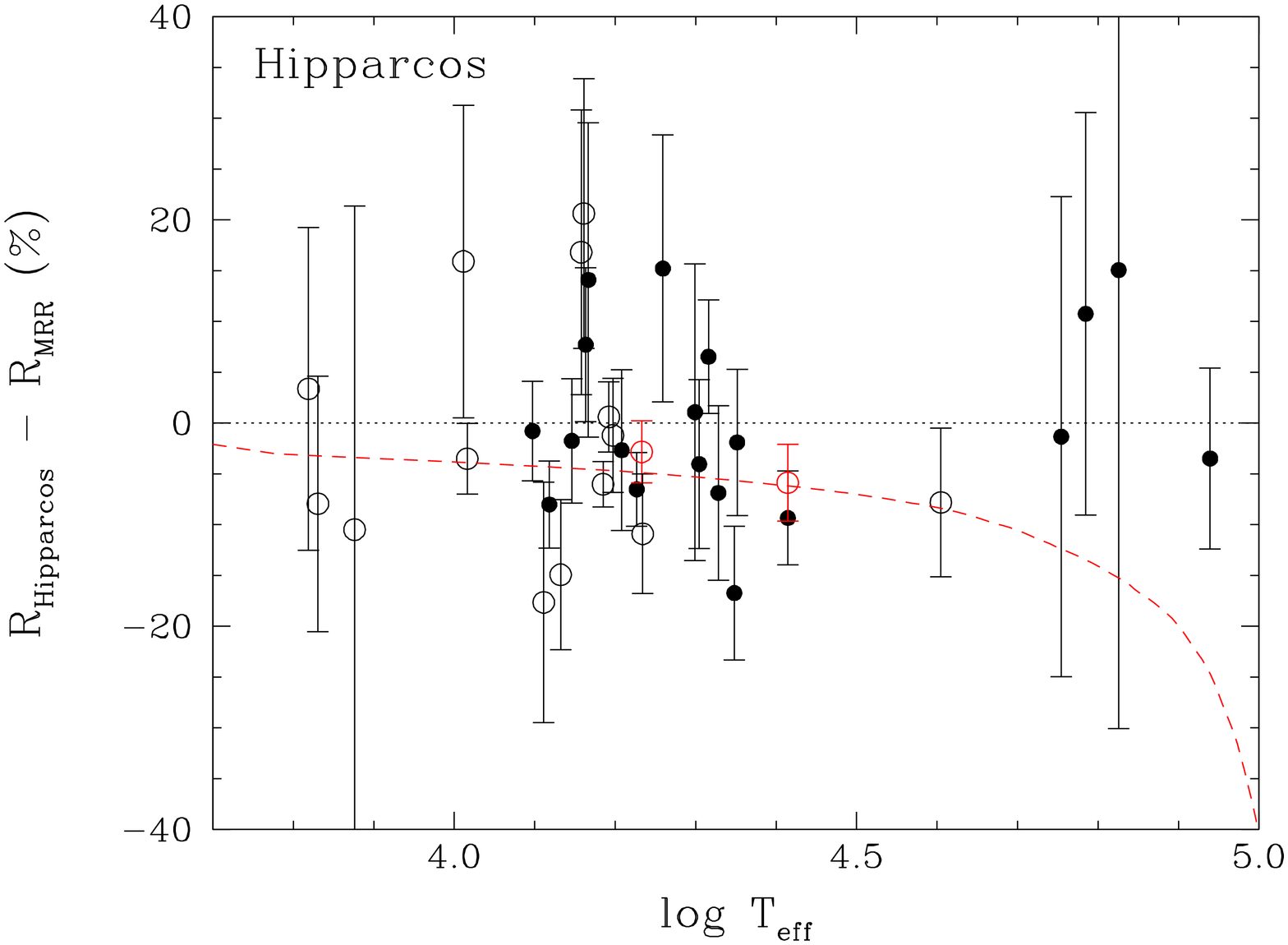}
  \caption[f2.eps]{{\it (Top:)} Differences (in \%) between observed {\it Gaia}
    DR1 radii $R_{\rm Gaia}$ (Eq.~\ref{eq2}) and predicted radii $R_{\rm
      MRR}$ drawn from the MRR with thick H-layers ($q_{\rm H} =
    10^{-4}$) as a function of $\log T_{\rm eff}$. Error bars for $\log T_{\rm
      eff}$ are omitted for clarity. Directly observed white dwarfs from
    Table~\ref{fg:t1} are represented by solid circles while wide binaries
    from Table~\ref{fg:t2} are illustrated by open circles. Numerical values
    are identified in Table~\ref{fg:t3}. The dotted line $\Delta R = 0$ is
    shown as a reference and the dashed red line is for a MRR relation with
    thin H-layers ($q_{\rm H} = 10^{-10}$) at 0.6 $M_{\odot}$. {\it (Bottom:)}
    Similar to the top panel but with pre-{\it Gaia} parallax measurements
    (mostly from {\it Hipparcos}) identified in Tables~\ref{fg:t1} and
    \ref{fg:t2}. We still rely on {\it Gaia} $G$ magnitudes when
    available. The benchmark cases 40 Eri B (cooler) and Sirius B (warmer) are
    shown in red.
    \label{fg:f3}}
\end{figure}

Fig.~\ref{fg:f3} compares $R_{\rm Gaia}$ (top panel) and $R_{\rm Hipparcos}$
(bottom panel) to $R_{\rm MRR}$. The dotted black line centered on zero
illustrates a perfect match between observations and theory for thick
H-layers, while the dashed red line shows the match to an illustrative
theoretical MRR with thin H-layers ($q_{\rm H} = 10^{-10}$) at 0.6
$M_{\odot}$. On average, the data agree with the theoretical MRR for thick
H-layers within 0.99$\sigma$ and 0.98$\sigma$ for {\it Gaia} DR1 and {\it
  Hipparcos}, respectively, and no significant systematic offset is observed
(neglecting the suspected double degenerates). The observed uncertainties for
both samples do not allow, however, for meaningful constraints on H envelope
masses. The error bars are only slightly smaller for the {\it Gaia} DR1 sample
compared to {\it Hipparcos}. There are two reasons for this behaviour. First
of all, most of the {\it Gaia} DR1 white dwarfs are companions to fairly
distant but bright primary stars with parallaxes. While the absolute parallax
error is on average 3 times smaller in {\it Gaia} DR1, the relative errors ($\sigma_{\pi}/\pi$)
are comparable with 5.05\% in {\it Gaia} DR1 and 7.06\% for pre-{\it Gaia}
measurements. Furthermore, the uncertainties from the atmospheric parameters
become the dominant contribution for the {\it Gaia} DR1 sample (see Section
4.2). The implications of these results are further discussed in Section~4.

\section{Discussion}

\subsection{Comparison with Other Empirical Mass-Radius Relations}

\begin{figure}
  \centering \includegraphics[scale=0.33,bb=30 80 592 722,angle=270]{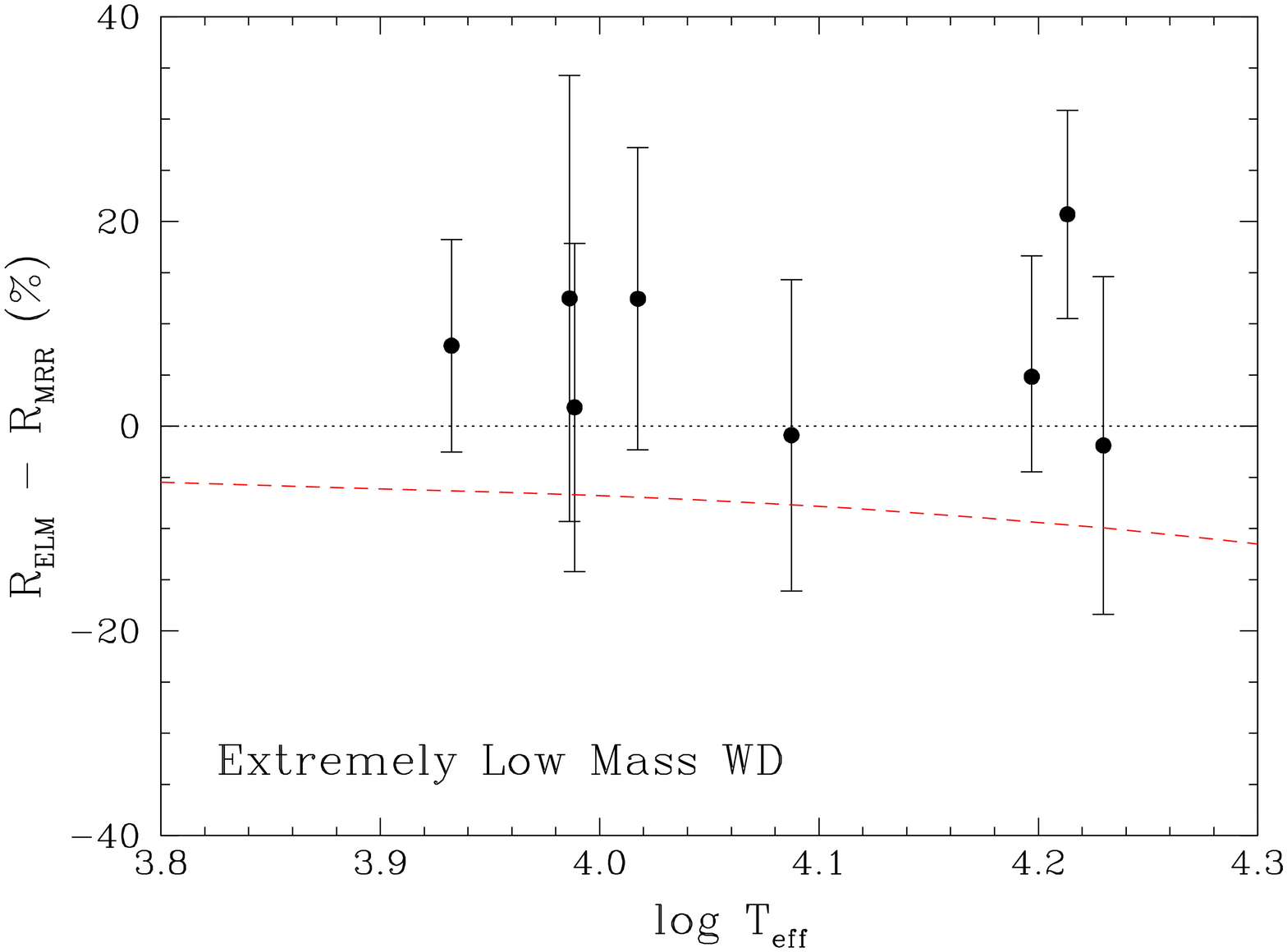}
  \centering \includegraphics[scale=0.33,bb=30 80 592 722,angle=270]{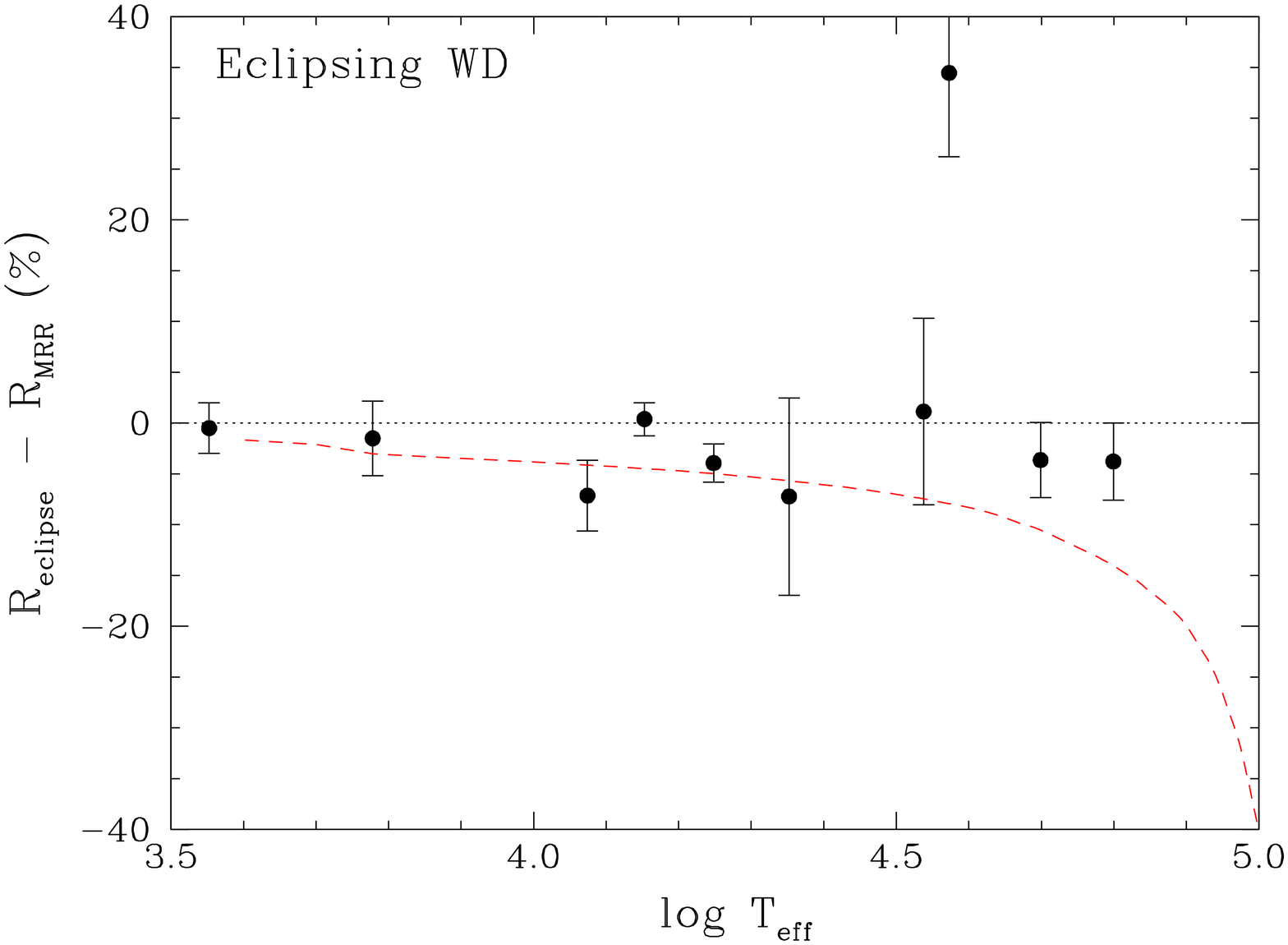}
  \caption[f4.eps]{{\it (Top:)} Differences (in \%) between observed radii
    $R_{\rm ELM}$ and predicted He-core radii $R_{\rm MRR}$ as a function of $\log
    T_{\rm eff}$ for the sample of He-core ELM white dwarfs from \citet{gianninas14}
    with 3D corrections from \citet{tremblay15}. 
    Error bars for $\log T_{\rm eff}$ are omitted for clarity and numerical values are presented in
    \citet{tremblay15}. The dotted line $\Delta R = 0$ is shown as a
    reference and the dashed red line is for a He-core MRR relation with thin H-layers
    at 0.3 $M_{\odot}$. {\it (Bottom:)} Differences between observed radii $R_{\rm
      eclipse}$ and predicted radii $R_{\rm MRR}$ for eclipsing binaries for
    which there is an independent derivation of both the mass and radius. The
    observed sample of both He- and C/O-core white dwarfs drawn from the literature is described in Table~\ref{fg:t4}.
    The dashed red line is for a MRR relation with thin H-layers at 0.6 $M_{\odot}$.
    \label{fg:f4}}
\end{figure}

Our results can be compared to two empirical MRRs not drawn from {\it
  Gaia} DR1. Fig.~\ref{fg:f4} (top panel) shows an independent analysis for
eclipsing and/or tidally distorted extremely low-mass (ELM) He-core white
dwarf systems that provide model-independent radii
\citep{hermes14,gianninas14}. The data are reproduced from table 7 of
\citet{tremblay15} where 3D model atmosphere corrections were applied.  The
theoretical radius $R_{\rm MRR}$ is taken from the spectroscopic atmospheric
parameters and the He-core MRR, similarly to our main analysis. The agreement
with the theoretical He-core MRR for thick H-layers is on average within error
bars. This result suggests that
the consistency between the theoretical MRR and spectroscopic atmospheric parameters
holds in the ELM regime as well.

Fig.~\ref{fg:f4} (bottom panel) also shows the results for eclipsing binaries
where masses and radii are both directly constrained from the eclipses 
and orbital parameters. The selected
systems from the literature and their parameters are identified in
Table~\ref{fg:t4}. In those cases, the theoretical
radius $R_{\rm MRR}$ is simply the dynamical mass processed through the
theoretical MRR for thick H-layers, hence the prediction is independent of the atmospheric parameters. 
The error bars are significantly smaller than those shown in Fig.~\ref{fg:f3} for {\it Gaia} DR1 and {\it Hipparcos}, leading
to a reduced y-axis scale in Fig.~\ref{fg:f4}. As discussed in
\citet{parsons16}, in most cases the observed radius is in agreement with the
theoretical MRR for thick H-layers. A mixture of He-cores ($M \leq 0.45$
$M_{\odot}$) and C/O-cores were employed given the masses of the white dwarfs
identified in Table~\ref{fg:t4}. SDSS~0857+0342 with 0.514 $M_{\odot}$ is the
one object in Fig.~\ref{fg:f4} that does not agree well with the C/O-core
MRR. \citet{parsons12a} have suggested that it might instead be a He-core
white dwarf. 

It may not be entirely surprising that none of these post-common envelope
systems are DB white dwarfs owing to the stellar wind of the companion. Very
few hydrogen deficient degenerates are known in post-common envelope systems \citep[see,
e.g.,][]{nagel06}. However, there is no evidence that the H envelope masses
are necessarily close to the maximum value of $q_{\rm H} \sim 10^{-4}$, and
the scatter observed in Fig.~\ref{fg:f4} could be due to these variations.
We remind the reader that H envelope mass determinations are model dependent
even for eclipsing binaries. 
The {\it Gaia} empirical MRR for single DA and DB white dwarfs could have more
objects with very thin H-layers, but there is no clear indication that the relation
would be significantly different. In particular, the results of
Fig.~\ref{fg:f4} for eclipsing binaries strongly suggest that theoretical
MRRs are in agreement with observations. The semi-empirical MRR for the 
{\it Gaia} DR1 sample in Fig.~\ref{fg:f3} supports this conclusion, but it also indicates 
that the spectroscopic atmospheric parameters are on average 
consistent with {\it Gaia} DR1 parallaxes. In future {\it Gaia} data releases, 
the results from eclipsing binaries may provide the key to disentangle a 
genuine observed signature of the white dwarf MRR from a systematic effect
from model atmospheres. 

Finally, we note that \citet{bergeron07} compared gravitational redshift
measurements with spectroscopically determined $\log g$ and a theoretical MRR,
but the comparison remained inconclusive because of the large uncertainties
associated with the redshift velocities.

\begin{table}
\scriptsize
\centering
\caption{Empirical Mass-Radius Relation from Eclipsing Binaries}
\setlength{\tabcolsep}{3.25pt}
\begin{tabular}{@{}cccccc}
\hline
\hline
\noalign{\smallskip}
Name & $M_{\rm eclipse}$ & $R_{\rm eclipse}$ & $R_{\rm MRR}$ & $T_{\rm eff}$ & Ref\\
     & [$M_{\odot}$]     & [$0.01R_{\odot}$] & [$0.01R_{\odot}$] & [K] & \\

 \noalign{\smallskip}
 \hline
 \noalign{\smallskip}
NN Ser            & 0.535 (0.012) & 2.08  (0.02)  & 2.16 (0.08)   & 63000 (3000)    & 1 \\
V471 Tau          & 0.840 (0.050) & 1.07  (0.07)  & 1.06 (0.07)   & 34500 (1000)    & 2 \\
SDSS J1210+3347   & 0.415 (0.010) & 1.59  (0.05)  & 1.61 (0.03)   & 6000  (200)     & 3 \\ 
SDSS J1212$-$0123 & 0.439 (0.002) & 1.68  (0.03)  & 1.75 (0.01)   & 17710 (40)      & 4 \\ 
GK Vir            & 0.562 (0.014) & 1.70  (0.03)  & 1.76 (0.06)   & 50000 (670)     & 4 \\
SDSS 0138$-$0016  & 0.529 (0.010) & 1.31  (0.03)  & 1.32 (0.01)   & 3570  (100)     & 5 \\ 
SDSS 0857+0342    & 0.514 (0.049) & 2.47  (0.08)  & 1.74 (0.15)   & 37400 (400)     & 6 \\ 
CSS 41177A        & 0.378 (0.023) & 2.224 (0.041) & 2.39 (0.22)   & 22500 (60)      & 7 \\ 
CSS 41177B        & 0.316 (0.011) & 2.066 (0.042) & 2.21 (0.06)   & 11860 (280)     & 7 \\
QS Vir            & 0.781 (0.013) & 1.068 (0.007) & 1.064 (0.016) & 14220 (350)     & 8 \\ 
 \noalign{\smallskip}
 \hline
 \noalign{\smallskip}
 \multicolumn{6}{@{}p{0.45\textwidth}@{}}{{\bf References.}
1) \citet{parsons10},
2) \citet{obrien01},
3) \citet{pyrzas12},
4) \citet{parsons12b},
5) \citet{parsons12c},
6) \citet{parsons12a},
7) \citet{bours15},
8) \citet{parsons16}.
\label{fg:t4}}
\end{tabular}
\end{table}

\subsection{Precision of the Atmospheric Parameters}

\begin{figure}
  \centering \includegraphics[scale=0.40,bb=180 190 362 582]{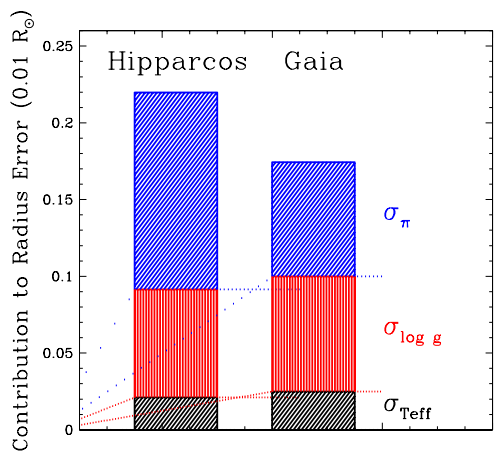}
  \caption[f2.eps]{Average error budget in the comparison of observed
    radii ($R_{\rm Gaia}$ or $R_{\rm Hipparcos}$) and predicted radii ($R_{\rm
      MRR}$) in Fig.~\ref{fg:f3}.  The different uncertainties are identified
    in the legend.
	\label{fg:f5}}
\end{figure}

The studies of \citet{vauclair97} and \citet{provencal98} have pioneered the
derivation of the semi-empirical MRR for white dwarfs using precise {\it
  Hipparcos} parallaxes. Our work with {\it Gaia} DR1 parallaxes is in
continuation of this goal. We remind the reader that such observed MRR is
still highly dependent on the white dwarf atmospheric parameters, hence model
atmospheres.  In previous studies, parallax errors were often dominant, but
with {\it Gaia} DR1 parallaxes, errors on spectroscopic atmospheric parameters
are becoming the most important. Fig.~\ref{fg:f5} illustrates the error budget
on $R_{\rm Gaia} - R_{\rm MRR}$ derived in Fig.~\ref{fg:f3} and demonstrates that
the uncertainties on $T_{\rm eff}$ and $\log g$ marginally dominate. The number and
precision of parallaxes will increase significantly with future {\it Gaia}
data releases. In particular, the individual parallaxes in DR2 will have significantly 
higher individual precision due to a longer measurement time (22 months instead of 11 months, 
which is already 36\% of the total mission time). Systematic errors are also expected
to decrease significantly resulting from a more sophisticated calibration, including 
a better definition of the line spread function, the application of a chromaticity correction, 
a more accurate calibration of the basic angle variation, and a calibration and correction of micro clanks. On the other hand, 
it is not expected that the precision on the atmospheric parameters will markedly improve anytime soon.

\begin{figure}
  \centering \includegraphics[scale=0.33,bb=30 50 592 762,angle=270]{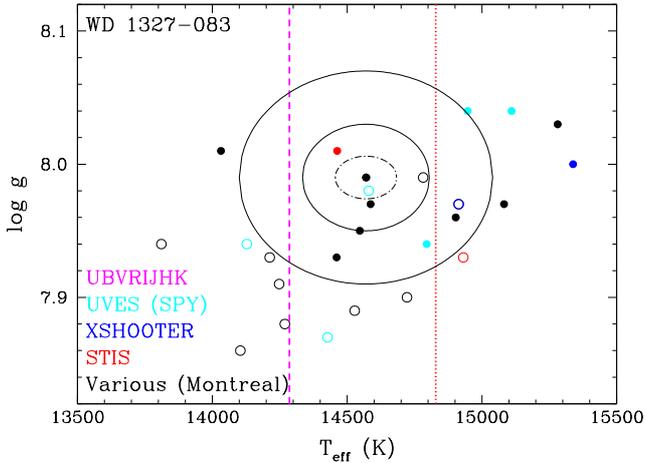}
  \caption[f6.eps]{Characterisation of the atmospheric parameters for WD~1327$-$083 
   using different observations and model atmospheres. The
    standard atmospheric parameters from \citet{gianninas11} used throughout
    this work are represented by their 1$\sigma$ and 2$\sigma$ error ellipses
    (solid black). The smaller formal $\chi^2$ error is represented by a dash-dotted
		ellipse. Different Balmer line solutions based on the same model atmospheres
    and fitting technique but alternative spectra are shown with solid
    circles. The alternative spectra are drawn from the Montreal group
    (black), the UVES instrument (SPY survey, cyan), X-SHOOTER (blue), and STIS
    spectrophotometry (red). We also show the alternative solutions employing the model
    atmospheres of \citet{koester10} with open circles. The formal
		$\chi^2$ error is very similar for all solutions.	Finally, we show our best
    fits of the continuum flux of STIS spectrophotometry (dotted red, see Fig.~\ref{fg:f7}) and UBVRIJHK
    photometry (dashed magenta, $\sigma_{\rm Teff} =$ 900~K). For photometric fits we have fixed the surface
    gravity at $\log g$ = 8.0.
    \label{fg:f6}}
\end{figure}

We propose that the bright and well-studied single DA white dwarfs in the {\it
  Hipparcos} sample, unfortunately largely missing from {\it Gaia} DR1, may be
used as a benchmark to understand the precision of the semi-empirical MRR of
future {\it Gaia} data releases. We will now assess the possibility of
improving the precision on the atmospheric parameters for these white dwarfs,
taking WD~1327$-$083 as an example. There are three steps in the Balmer line
fitting procedure that could introduce errors; uncertainties in the
spectroscopic data, issues with the fitting procedure, and inaccuracies in the
model atmospheres. To illustrate this, we have derived the atmospheric
parameters of WD~1327$-$083 using a number of observations and methods. In
Fig.~\ref{fg:f6} we display the published \citet{gianninas11} atmospheric
parameters based on one spectrum. The formal $\chi^2$ uncertainty
is represented by the smaller dash-dotted ellipse. We remind the reader that the error bars from
\citet{gianninas11} combine in quadrature this formal $\chi^2$ error and a fixed
external error of 1.2\% in $T_{\rm eff}$ and 0.038 dex in $\log g$, resulting in
the corresponding 1$\sigma$ and 2$\sigma$ error ellipses shown in Fig.~\ref{fg:f6}.

\begin{figure}
  \centering \includegraphics[scale=0.33,bb=30 50 592
  762,angle=270]{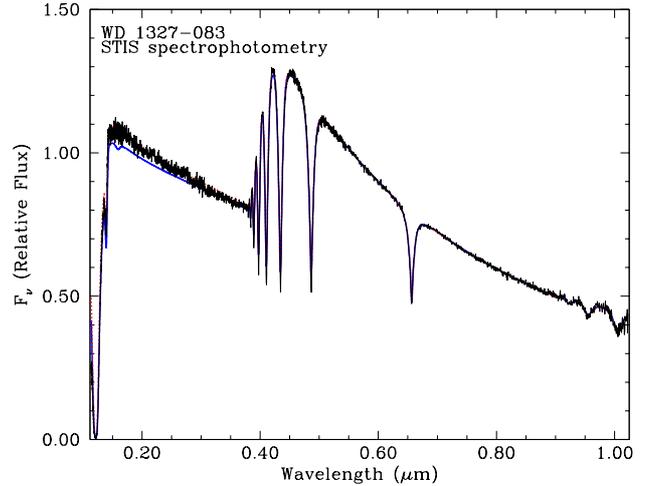}
  \caption[f7.eps]{STIS spectrophotometric observations of WD~1327$-$083 as 
a function of wavelength. The predicted flux from the model atmospheres of
\citet{tremblay11} using the atmospheric parameters of \citet{gianninas11}
is shown in blue (solid, $T_{\rm eff} = 14,570$ K, $\log g$ = 7.99), and the 
best fit is shown in red (dotted, $T_{\rm eff} = 14,830$ K with $\log g$ fixed at 8.0),
which is almost coincident with the observations on this scale.
\label{fg:f7}}
\end{figure}

First of all, we rely on 12 alternative spectra for WD~1327$-$083. 
These are all high signal-to-noise (S/N $> 50$) observations that were
fitted with the same model atmospheres \citep{tremblay11} and the same fitting
code as in \citet{gianninas11}. In all cases the formal $\chi^2$ error
is very similar to the one illustrated in Fig.~\ref{fg:f6} for the spectrum
selected in \citet{gianninas11}. We employ 7 spectra taken by the Montreal group
from different sites (black filled points in Fig.~\ref{fg:f6}) in addition to the one selected in
\citet{gianninas11}. We also rely on 3 UVES/VLT spectra taken as part of the SPY survey 
\citep{koester09}, shown with cyan filled circles in Fig.~\ref{fg:f6}. 
Additionally, new observations were secured. The first one is a
high S/N X-SHOOTER/VLT spectrum taken on programme 097.D-0424(A). The
Balmer lines suggest a significantly warmer temperature (blue filled circle)
than the average in Fig.~\ref{fg:f6}. However, the calibrated spectra
show a smaller than predicted flux in the blue, suggesting the offset could be
caused by slit losses during the observations. Finally, we have recently
obtained STIS spectrophotometry for WD~1327$-$083 under {\it Hubble Space
  Telescope} program 14213 as shown in Fig.~\ref{fg:f7}.  The Balmer lines
were fitted and a solution (red filled circle in Fig.~\ref{fg:f6}) very
similar to that of \citet{gianninas11} was obtained.

The atmospheric parameters in Fig.~\ref{fg:f6}, determined from different
spectroscopic data, show a relatively large scatter that is significantly higher than the
$\chi^2$ error, confirming that external errors from the data reduction must
be accounted for. The scatter appears slightly larger than the systematic uncertainty
estimated by \citet{LBH} and \citet{gianninas11} from a similar procedure. 
However, one could argue that some of the observations selected in 
this work should have a lower weight in the average since they show minor
deficiencies in their instrumental setup or flux calibration. 

The STIS spectrophotometry, which is calibrated using the three hot ($T_{\rm
  eff} > 30,000$ K) white dwarfs GD 71, GD 153, and G191$-$B2B
\citep{bohlin14}, also permits the determination of the atmospheric parameters
based on the continuum flux. The surface gravity was fixed at $\log g = 8.0$
since the sensitivity of the continuum flux to this parameter is much smaller
than the sensitivity to $T_{\rm eff}$. The blue wing and central portion of Ly
$\alpha$ were removed from the fit because the observed flux is very small in
this region. Fig.~\ref{fg:f7} shows our best-fit model (red) compared to the
solution using the $T_{\rm eff}$ value from \citet{gianninas11} in blue. The
solution is clearly driven by the UV flux, and a $T_{\rm eff}$ value of
14,830~K, about 250~K larger than that of \citet{gianninas11}, is required to
fit the observations. The STIS photometric solution is added to
Fig.~\ref{fg:f6} (dotted red line). It is reassuring that there is a good consistency
between STIS spectrophotometry and white dwarf atmospheric parameters both for
current hotter flux standards and this cooler object.  A full
discussion about using this white dwarf as a STIS spectrophotometric standard
will be reported elsewhere. As an independent test, we have also used UBVRIJHK
data drawn from \citet{koen10} and 2MASS \citep{2mass} to fit a temperature of
14,285 $\pm$ 900~K. The large error is due to the fact that this photometric
data set does not include the UV which is the most sensitive to $T_{\rm
  eff}$. We refrain from using the {\it GALEX} FUV and NUV fluxes since there
is a significant systematic offset between observed and synthetic fluxes in
the magnitude range of WD~1327$-$083 \citep{galex}. The results are reported
in Fig.~\ref{fg:f6} (dashed magenta), though because of the large error, the
UBVRIJHK $T_{\rm eff}$ value is fully consistent with the STIS
spectrophotometry.

Finally, we have performed the same analysis but using instead the model
atmospheres of \citet{koester10} including the Stark broadening profiles of
\citet{tremblay09}.  The results are shown in Fig.~\ref{fg:f6} with open
circles for fits of the Balmer lines. The mean $T_{\rm eff}$ value is shifted
by $-$295~K and the mean $\log g$ value by $-$0.06 dex, which is in both cases
slightly larger than the published error bars. In the case of the STIS and
UBVRIJHK photometric fits, we find essentially the same $T_{\rm eff}$ values
with both grids of models.

Fig. \ref{fg:f6} demonstrates that for the particular case of WD~1327$-$083,
the 1$\sigma$ error bars from \citet{gianninas11} are a reasonable but likely
optimistic estimate of the $T_{\rm eff}$-$\log g$ uncertainties. It is perhaps
not surprising since they did not consider alternative model grids or
photometric solutions in their uncertainties. We have not explicitly
considered the effect of the fitting techniques, which would increase even
more the scatter between the different solutions. However, changing the
fitting method would not provide a fully independent diagnostic since it is
influenced by both the data reduction and systematic uncertainties in the
model atmosphere grids.

It is outside the scope of this work to review the differences between the
model grids or to re-observe spectroscopically all white dwarfs for which we currently have
parallaxes. Nevertheless, we suggest that this should be done ahead of {\it
  Gaia} DR2 for a benchmark sample of bright white dwarfs. We can nevertheless
make a few additional observations. If we allow the uncertainties on the
atmospheric parameters to increase by a very conservative factor of two
following our discussion above, 21/26 {\it Gaia} DR1 white dwarfs
  agree within error bars with thick H-layers while 22/26 are consistent with thin H-layers. These
  results suggest that given the precision on the atmospheric parameters, the
  theoretical MRR is entirely consistent with the observations. Furthermore,
  the distinction between thin and thick H-layers for {\it Gaia} DR1 white
  dwarfs is still out of reach, as it was the case for {\it Hipparcos}.

\section{CONCLUSIONS}

The {\it Gaia} DR1 sample of parallaxes was presented for 6 directly
  observed white dwarfs and 46 members of wide binaries.  By combining this
  data set with spectroscopic atmospheric parameters, we have derived the semi-empirical
  MRR relation for white dwarfs. We find that, on average, there is a good
  agreement between {\it Gaia} parallaxes, published $T_{\rm eff}$ and $\log
  g$, and theoretical MRRs. It is not possible, however, to conclude that both
  the model atmospheres and interior models are individually consistent with
  observations. There are other combinations of $T_{\rm eff}$, $\log g$, and H
  envelope masses that could agree with {\it Gaia} DR1 parallaxes.  However,
  the good agreement between observed and predicted radii for eclipsing
  binaries, which are insensitive to model atmospheres, suggest that both the
  atmospheric parameters and theoretical MRRs are consistent with {\it
    Gaia} DR1. 

  Starting with {\it Gaia} DR2, it will be feasible to derive the
  semi-empirical MRR for thousands of white dwarfs. Assuming systematic parallax errors
	will be significantly reduced, it will be possible
  to take advantage of large number statistics and compute a precise
  offset between the observed and predicted MRRs for $T_{\rm
    eff}$, mass, and spectral type bins. Alternatively, since the mass and radius are derived quantities, 
  the parallax distances could be directly compared to predicted spectroscopic distances
  \citep{holberg08}. However, it may be difficult to interpret the results in
  terms of the precision of the model atmospheres and evolutionary
  models. Independent constraints from eclipsing binaries, as well as a more
  careful assessment of the error bars for bright and well known white dwarfs,
  may still be necessary to fully understand {\it Gaia} data.

\section*{Acknowledgements}

This project has received funding from the European Research Council (ERC)
under the European Union's Horizon 2020 research and innovation programme
(grant agreements No 677706 - WD3D and No 320964 - WDTracer). TRM is grateful
to the Science and Technology Facilities Council for financial support in the
form of grant No ST/L000733. This work has made use of data from the ESA
space mission Gaia, processed by the Gaia Data Processing and Analysis
Consortium (DPAC). Funding for the DPAC has been provided by national
institutions, in particular the institutions participating in the Gaia
Multilateral Agreement. The Gaia mission website is
\url{http://www.cosmos.esa.int/gaia}. This work is based on observations made with
the NASA/ESA Hubble Space Telescope, obtained at the Space Telescope Science
Institute, which is operated by the Association of Universities for Research
in Astronomy, Inc., under NASA contract NAS 5-26555. These observations are
associated with programme \#14213. This paper is based on observations made
with ESO Telescopes under programme ID 097.D-0424(A).

\bsp % typesetting comment
\label{lastpage}
\end{document}